\documentclass[11pt,a4paper,english,twoside]{article}

\usepackage{a4wide}
\usepackage{amssymb, amsmath}
\usepackage{graphicx}
\usepackage[all]{xy}
\usepackage{hyperref}
\usepackage[T1]{fontenc}
\hypersetup{linktocpage}
\usepackage{enumerate}
\usepackage{xcolor}
\usepackage{lmodern}
\usepackage{dsfont}
\usepackage{empheq}
\usepackage{cite}
\usepackage[utf8]{inputenc}
\usepackage{wasysym}

\newcommand{\beq}{\begin{equation}}
\newcommand{\eeq}{\end{equation}}
\def\bea#1\eea{\begin{align}#1\end{align}}
\def\beal#1\eeal{\begin{subequations}\begin{align}#1\end{align}\end{subequations}}
\newcommand{\nn}{\nonumber}
\newcommand{\w}{\wedge}

\renewcommand{\i}{\ensuremath{\textnormal{i}}}

\def\del {\partial}
\def\d {{\rm d}}
\def\mmm {\mathcal{M}}

\begin{document}
\numberwithin{equation}{section}

\begin{titlepage}

\begin{center}

\phantom{DRAFT}

\vspace{3.1cm}

{\LARGE \bf{Signatures of extra dimensions in gravitational waves}}\\

\vspace{2 cm} {\Large David Andriot and Gustavo Lucena G\'{o}mez}\\
 \vspace{0.9 cm} {\small\slshape Max-Planck-Institut f\"ur Gravitationsphysik, Albert-Einstein-Institut,\\Am M\"uhlenberg 1, 14467 Potsdam-Golm, Germany}\\
\vspace{0.5cm} {\upshape\ttfamily andriotphysics@gmail.com, glucenag@aei.mpg.de}\\

\vspace{3cm}

{\bf Abstract}
\end{center}

\begin{quotation}
\noindent
Considering gravitational waves propagating on the most general $4+N$-dimensional space-time, we investigate the effects due to the $N$ extra dimensions on the four-dimensional waves. All wave equations are derived in general and discussed. On Minkowski$_4$ times an arbitrary Ricci-flat compact manifold, we find: a massless wave with an additional polarization, the breathing mode, and extra waves with high frequencies fixed by Kaluza--Klein masses. We discuss whether these two effects could be observed.
\end{quotation}

\end{titlepage}

\newpage

\tableofcontents

\vspace*{10pt}

\section{Introduction}

The direct observations of gravitational waves emitted by black hole mergers, realised by the LIGO and Virgo Scientific Collaboration \cite{Abbott:2016blz,Abbott:2016nmj}, are impressive experimental results and groundbreaking scientific achievements. They provide physicists with a new observational tool, allowing them to probe nature and test theories in completely innovative manners. Constraints on various models beyond four-dimensional General Relativity have already been obtained from these observations, for instance constraints on alternative theories of gravity \cite{Konoplya:2016pmh,Chesler:2017khz}, modified dispersion relations \cite{Arzano:2016twc}, quantum gravity effects \cite{Calmet:2016sba, Barcelo:2017lnx}, non-commutative geometry \cite{Kobakhidze:2016cqh}, exotic compact objects \cite{Cardoso:2016oxy}, microscopic description of black holes~\cite{Brustein:2017koc}, etc. A recent review on such constraints can be found in \cite{Yunes:2016jcc}. In the present paper, we study the consequences of putative extra dimensions on four-dimensional gravitational waves, and whether related effects could be detected.

\subsection*{Background and state of the art}\addcontentsline{toc}{subsubsection}{Background and state of the art}

Considering spatial dimensions in addition to our four-dimensional space-time is a common idea when going beyond standard particle physics, gravity or cosmology. Such extra dimensions appear in models ranging from phenomenological or bottom-up approaches, to quantum gravity theories such as string theories, and their low energy realisations as supergravities. The former include models with large extra dimensions \cite{Antoniadis:1990ew}, in particular so-called ADD models \cite{ArkaniHamed:1998rs, Antoniadis:1998ig}, Randall--Sundrum models RS1 \cite{Randall:1999ee} and RS2 \cite{Randall:1999vf}, and models with Universal Extra Dimensions (UED) \cite{Appelquist:2000nn}. A first distinction between these models is about matter and interactions: while gravity is present in all dimensions, matter and other gauge interactions can be restricted to a subset of dimensions, referred to as the brane, such as for example the four-dimensional space-time. This holds for ADD and RS models, but not for UED. Such a restriction provides an explanation to the hierarchy between the Planck mass and other energy scales such as the electro-weak scale, either thanks to large extra dimensions (ADD) or to a warp factor (RS). Another distinction is on the number and compactness of extra dimensions: there can be one (RS1) or several (UED, ADD) compact extra dimensions, or one extended extra dimension (RS2). Several of these features are present in ten-dimensional type II supergravities and their compactifications, the low-energy effective theories of type II string theories (reviews can be found e.g.~in~\cite{Grana:2005jc, Blumenhagen:2006ci}). These theories feature six extra dimensions, gathered as compact manifolds: those can be Ricci flat as Calabi--Yau manifolds, or curved as e.g.~Lie group manifolds \cite{Andriot:2010ju, Danielsson:2011au}. They also admit branes localizing non-abelian gauge interactions, warp factors, etc. A short review on all these models and their connections can be found in \cite{Quiros:2006my}, see also~\cite{Giddings:2016sfr}. Constraints exist for each of them, but possibilities on the number, shape, and size, of extra dimensions remain very open. As a consequence, our paper aims at finding general, if possible model-independent signatures of extra dimensions in four-dimensional gravitational waves.

Considering gravitational waves in a space-time of dimension $D=4+N$, with $N$ extra dimensions, is not a new idea. An important load of work has been devoted to studying the emission of such waves from black holes in $D$ dimensions; see e.g. the review of black holes in higher dimensions \cite{Emparan:2008eg}. A seminal paper on this topic is \cite{Cardoso:2002pa}, considering colliding particles (possibly black holes) as sources and obtaining the quadrupole formula. Subsequent work, e.g.~\cite{Berti:2003si, Cardoso:2008gn}, was motivated in part by the possibility of black hole creation at the LHC, and their stability under gravitational perturbations and radiation. Related work, e.g.~\cite{Ishibashi:2003ap, Konoplya:2003dd, Creek:2006ia, LopezOrtega:2006my}, focused on computing quasinormal modes of a $D$-dimensional black hole. The methods developed include numerical approaches, see e.g.~\cite{Cook:2016qnt} and references therein. A review on this topic can be found in \cite{Kanti:2004nr}. From our perspective, there are two drawbacks to those studies. First, the sources considered for the gravitational waves are very specific, while many others could exist (see e.g.~\cite{Cutler:2002me, Cai:2017cbj} for reviews). Second, the background metric away from the black hole is the Minkowski flat one, thus describing a $D$-dimensional Minkowski space-time, or at most $N$ extra circles (equivalently a flat $N$-torus $T^N$). This is a strong restriction on the extra dimensions.

Other works have focused more directly on potentially detectable effects of extra dimensions on gravitational waves, often in a specific model or background. By studying the signal's Green function, a tail effect was pointed out in \cite{Barvinsky:2003jf, Cardoso:2003jf} for a one-circle extra dimension, or on a $D$-dimensional de Sitter or FLRW background in \cite{Chu:2016qxp, Chu:2016ngc}. A small correction to the waveform was also obtained in \cite{Qiang:2017luz} with fairly general compact extra dimensions and a four-dimensional source. An interesting polarization effect was obtained in \cite{Alesci:2005rb} for a single circle extra dimension, with some Ansatz for the fields. Another interesting idea is that gravitational waves at high frequencies can provide a hint at extra dimensions and high energy physics. It is present in various papers considering RS models, first in \cite{Hogan:2000is, Hogan:2000aa} with waves emitted in the early universe, and then in \cite{Clarkson:2006pq} where discrete frequencies of some gravitational radiation are related to Kaluza--Klein modes of a specific RS model. A similar idea and model, but with a continuous spectrum, can be found in \cite{Inoue:2003di}. Further work includes effects due to cosmic strings \cite{DePies:2007bm, OCallaghan:2010rlo}, braneworlds \cite{deRham:2005jan}, gravitational leakage \cite{Deffayet:2007kf}, or inflation considerations \cite{GomezMartnez:2007aaw}. We will come back to some of these effects.

\subsection*{Description of the work}\addcontentsline{toc}{subsubsection}{Description of the work}

In this paper, we take a more general approach: in short, sources are left unspecified, and extra dimensions are not restricted, except through their background equations. As a result, we identify and discuss general effects of extra dimensions on four-dimensional gravitational waves, which could in principle be observed. As mentioned above, extra dimensions appear in a wide variety of models, and their number, size or geometry is so far not sharply constrained. Given this set of possibilities, we initially refrain from specifying this geometry and consider a generic background. Similarly, the four-dimensional space-time is at first not restricted to be Minkowski: we allow for general geometries, in particular de Sitter, which could be useful for primordial gravitational waves, or anti-de Sitter, of possible interest for holographic applications. When analysing in detail the effects of the extra dimensions, we will nevertheless approximate to a four-dimensional Minkowski space-time, better suited for the comparison to currently observed gravitational waves, times a Ricci flat $N$-dimensional compact manifold. Those extra dimensions, satisfying the background equations, are otherwise left unconstrained: there is for instance a huge, if not infinite, number of Calabi--Yau manifolds which would suit. In any case, the formulation developed would allow to consider more general geometries.

We also do not specify the source of gravitational waves: as mentioned, many different sources are possible, and their physics, especially in the hypothetical extra dimensions, is not known in general. Thus we do not consider the source and emission process: in the wave equations, there will not be any source term. Rather, information on the emission is taken as initial data of the waves, and we study the propagation (and detection) of gravitational waves in an empty $D$-dimensional space-time. This is a sensible approximation insofar as observed gravitational waves are believed to be so-called pristine probes of the emitted signals, because their interaction with matter is extremely weak. Eventually, we make qualitative predictions regarding the effects of extra dimensions on the four-dimensional wave, which could in principle be observed.\\

Our starting point is $D$-dimensional General Relativity with a cosmological constant; further contributions from other ingredients are considered in Appendix~\ref{ap:gen}. From there, we derive the equation of motion for the linear metric fluctuation $h_{MN}$ on a generic background (instead of the standard Minkowski flat metric), i.e.~the general $D$-dimensional gravitational wave equation. We then split space-time into $4$ and $N$ dimensions, and deduce the wave equations for the three types of components, namely the four-dimensional wave $h_{\mu\nu}$, the vectors $h_{\mu n}$ and the scalars $h_{mn}$, with ${}_{\mu}$ (respectively ${}_m$) being the four- (respectively $N$-) dimensional indices. We finally study the equations describing four-dimensional gravitational waves, and look for deviations from standard equations, i.e.~for an effect due to the extra dimensions.

In general there can be three types of effects, given the differences with the standard four-dimensional analysis. Starting with $h_{MN}(X^P)$ and focusing on $h_{\mu\nu}(x^{\pi})$, there are two differences: first, there are other components $h_{\mu n}$ and $h_{mn}$, and second, there are extra coordinates that we denote $y^m$. The first type of effect is due to a coupling between the various components, i.e.~$h_{\mu n}$ or $h_{mn}$ may enter the equations describing the four-dimensional gravitational waves. The second type of effect, due to the $y^m$, is that instead of a single four-dimensional wave, one gets a Kaluza--Klein tower of modes $h_{\mu\nu}^{{\bf k}}(x^{\pi})$, possibly coupled to each other. One of these modes is a priori massless, the others being massive. In different contexts, e.g.~missing energy at the LHC, the latter would correspond to so-called Kaluza--Klein gravitons \cite{Agashe:2007zd, Giddings:2016sfr}. However in the present work we consider them only classically, interpreting them as extra contributions to the four-dimensional gravitational wave. The third type of effect comes from the non-triviality background: for example, having a warp factor could lead to differences with respect to the standard situation with four-dimensional Minkowski. The focus of the present work is to study realisations of these three types of effects.\\

More concretely, with the four-dimensional space-time being Minkowski, times an arbitrary Ricci-flat compact manifold of dimension $N$, we find the following two signatures of extra dimensions on four-dimensional gravitational waves:
\begin{description}
	\item[1. Breathing mode:] Due to the extra scalars, a massless breathing mode is present in general, in addition to the usual cross and plus polarizations of the gravitational wave. It is characterized by a homogeneous deformation of the two transverse directions.
	\item[2. High-frequency signals:] Extra four-dimensional signals, verifying a massive dispersion relation, add up to the standard massless gravitational wave. They are characterized by a discrete set of higher frequencies, fixed by the Kaluza--Klein masses.
\end{description}
Along the way we derive the equations of motion for the three types of components in full generality, allowing for an arbitrary background geometry with a warp factor. Although we do not make a concrete prediction in the most generic case of a non-constant warp factor, we comment on the general impact of the latter on the propagation of four-dimensional waves. A more detailed summary of our results is relegated to Section \ref{sec:ccl}, together with a discussion on the observability of the above predictions. \\

The paper is organized as follows. In Section~\ref{sec:firstderiv} we derive the linearized Einstein equations with cosmological constant $\Lambda_D$ on the most general space-time of dimension $D=4+N$ with non-trivial warp factor. We first obtain the $D$-dimensional equation in the de Donder gauge in Subsection~\ref{sec:Ddim}, and then split both the equation and the gauge condition according to the $4+N$-dimensional structure of space-time in Subsection~\ref{sec:split}. An analogous derivation in the Transverse-Traceless gauge is performed in Appendix \ref{ap:TT}. In Section~\ref{sec:constantwarp} we restrict to the case of a Minkowski$_4$\,$\times$\,$\mathcal{M}$ space-time with an arbitrary compact and Ricci-flat manifold $\mathcal{M}$. We first study the equation of motion and gauge condition for the massless wave in Subsection~\ref{sec:massless}, where we unveil the presence of an additional breathing mode on top of the two usual polarizations of General Relativity. We then focus on the extra four-dimensional waves in Subsection~\ref{subsec:massive}, where we find they have all six polarization modes turned on with high frequencies related to Kaluza--Klein masses. Finally, Section~\ref{sec:ccl} contains a summary of our results as well as an extended discussion on the observability of the two above effects. Equations in the more involved cases of a non-constant warp factor, or additional content in the $D$-dimensional Lagrangian, are discussed in Appendix \ref{ap:nonconstantwarp}, and Appendix \ref{ap:gen}, respectively.\\

\noindent {\bf Note:} With respect to previous versions of this paper, a small computational mistake has been corrected in equation \eqref{boxh4d}, and consequently in equations \eqref{eqgen4d} and \eqref{eqgen4dTT}. This has no impact on the main analysis and results of the paper. However, Appendix \ref{ap:nonconstantwarp} got important modifications, starting with \eqref{eqh}, since the correction simplifies equations studied there. The resulting equations reproduce those obtained in previous works in the literature.

\section{Gravitational waves in a \texorpdfstring{$D$}{D}-dimensional space-time}\label{sec:firstderiv}

We start with General Relativity in dimension~$D$, derive the linearized equations of motion, and then split them according to a $4+N$-dimensional space-time.

\subsection{Gravitational-wave equation in \texorpdfstring{$D$}{D} dimensions}\label{sec:Ddim}

We derive the equation describing the propagation of gravitational waves in an empty $D$-dimensional space-time. To that end, we consider the action for General Relativity in dimension $D\geq 4$, with a cosmological constant $\Lambda_D$,
\beq
S= \frac{1}{2\kappa_D} \int \d^{D} x \, \sqrt{|g_{D}|}\ \left( {\cal R}_{D} - 2 \Lambda_D   \right)\ , \label{action}
\eeq
where $|g_{D}|$ denotes the absolute value of the determinant of the $D$-dimensional metric, with components $g_{D\ MN}$, ${\cal R}_{D}$ denotes the corresponding Ricci scalar, ${\cal R}_{MN}$ the corresponding Ricci tensor, and $\kappa_D$ is a constant.\footnote{The equations derived here, in particular the $D$-dimensional wave equation on a generic background, its decomposition on the various components, and further the Kaluza--Klein modes, may have already appeared in the context of dimensional reductions of supergravities, at least in some related forms. The reason is that formally, the same objects are considered. For instance, Section 5.1 of the review \cite{Duff:1986hr} contains a $D$-dimensional linearized Einstein equation, further decomposed into components, with however a different background (four-dimensional anti-de Sitter, constant warp factor) and different gauge fixings. Section 5.1 of the review \cite{Bailin:1987jd} gives the wave equation of four-dimensional Kaluza--Klein modes, with the background being Minkowski times a circle, and a specific gauge fixing. The focus of the present work is nevertheless different: while supergravity dimensional reductions typically consider particular backgrounds with field Ans\"atze and make related rearrangements of the equations, we remain very general regarding the background and gauge conditions, trying to capture all possible effects. In addition, we aim at interpretations in terms of (observable) four-dimensional gravitational waves, which is a very different perspective. The supergravity literature remains certainly useful in our context, at least on technical aspects.} Even for a propagation in an empty four-dimensional space-time, models with extra dimensions usually contain more terms in their action, given e.g.~by gauge fields and fluxes, branes, etc. A general Lagrangian accounting for such terms is considered in Appendix \ref{ap:gen}, where the following derivation of the gravitational-wave equation is extended. For now, we restrict ourselves to \eqref{action}, and derive the following Einstein equation:
\beq
{\cal R}_{MN} - \frac{g_{D\ MN}}{2}\left({\cal R}_D - 2\Lambda_D \right) = 0 \ . \label{EinsteinD}
\eeq
Using its trace
\beq
{\cal R}_D = \frac{2D}{D-2}\, \Lambda_D \ , \label{Einsteintrace}
\eeq
it can be rewritten as the following (trace-inversed) Einstein equation:
\beq
{\cal R}_{MN} - \frac{2\, \Lambda_D}{D-2}\, g_{D\ MN}= 0 \ .\label{Einstein}
\eeq

To describe gravitational waves, we now decompose the $D$-dimensional metric $g_{D\ MN}$ as the sum of a generic background $g_{MN}$ and a fluctuation $h_{MN}$,
\beq
g_{D\ MN} = g_{MN} + h_{MN} \ ,\label{gDdecompo}
\eeq
which enjoys a gauge transformation given by linearized diffeomorphisms,
\begin{equation}
\label{gaugevariationgeneric}
	h_{MN} \rightarrow h'_{MN} = h_{MN} + \delta h_{MN}\,, \qquad \delta h_{MN} = 2\nabla_{(M} \xi_{N)}\,,
\end{equation}
where $\xi_{N}$ is the infinitesimal gauge parameter. We will develop equations at linear order in $h$. We denote with ${}^{(0)}$ and ${}^{(1)}$ quantities at zeroth order (i.e.~background) and first order in $h$. The Einstein equation \eqref{Einstein} splits accordingly:
\bea
& {\cal R}^{(0)}_{MN} - \frac{2\, \Lambda_D}{D-2}\, g_{MN} = 0 \ ,\label{Einstein0}\\
& {\cal R}^{(1)}_{MN} - \frac{2\, \Lambda_D}{D-2}\, h_{MN} = 0 \ ,\label{Einstein1}
\eea
where we recall that $\Lambda_D$ is constant, and thus entirely captured by the background. Further, one verifies
\beq
(g + h)^{(0)\, MN}= g^{MN}, \ (g + h)^{(1)\, MN}= - g^{MP} h_{PQ} g^{QN} \ .
\eeq
We now need to compute ${\cal R}^{(1)}_{MN}$. The definitions of the relevant geometric quantities (with Levi--Civita connection) are given e.g.~in Appendix A of \cite{Andriot:2013xca}.

First, one can show for the connection coefficient that
\beq
\Gamma^M_{NP}= \Gamma^{(0)}{}^M_{NP} + \frac{1}{2} g^{MQ} \left(\nabla^{(0)}_N h_{QP} + \nabla^{(0)}_P h_{QN} - \nabla^{(0)}_Q h_{NP} \right) \ ,
\eeq
with the standard definition $\Gamma^{(0)}{}^M_{NP}= \tfrac{1}{2} g^{MQ} \left(\del_N g_{QP} + \del_P g_{QN} - \del_Q g_{NP} \right)$. Denoting the trace $h_D= h_{QP}g^{PQ}$, the Ricci tensor is then given by
\beq
{\cal R}_{MN}= {\cal R}^{(0)}_{MN} - \frac{1}{2} \nabla^{(0)}_P(g^{PQ} \nabla^{(0)}_Q h_{MN} ) + \nabla^{(0)}_P(g^{PQ} \nabla^{(0)}_{(M} h_{N)Q} ) - \frac{1}{2} \nabla^{(0)}_N \nabla^{(0)}_M h_D  \ .\label{Ricci}
\eeq
In view of a gauge fixing, we now make the following quantity appear:
\beq
{\cal G}_N \equiv \nabla^{(0)}_P g^{PQ} h_{QN} - \frac{1}{2} \nabla^{(0)}_N h_D \ . \label{Donder}
\eeq
This amounts to commuting covariant derivatives. Acting on a scalar they commute, while
\beq
\nabla^{(0)}_P \nabla^{(0)}_M g^{PQ} h_{QN} = \nabla^{(0)}_M \nabla^{(0)}_P g^{PQ} h_{QN} + {\cal R}^{(0)}_{MP} g^{PQ} h_{QN} + g_{NS} {\cal R}^{(0)}{}^S{}_{RPM} g^{PQ} h_{QU} g^{UR} \ .
\eeq
We deduce
\beq
{\cal R}_{MN}^{(1)}=  - \frac{1}{2} \square_D^{(0)}\, h_{MN} + g^{PQ} h_{Q(N} {\cal R}^{(0)}_{M)P}  + {\cal R}^{(0)}{}^S{}_{MNP}\, g^{PQ} h_{QS} + \nabla^{(0)}_{(M} {\cal G}_{N)} \ ,\label{Ricci2}
\eeq
with $\square_D^{(0)}= g^{PQ} \nabla^{(0)}_P \nabla^{(0)}_Q$, and ${\cal R}^{(0)}{}^S{}_{RP(M}  g_{N)S} g^{PQ} h_{QU} g^{UR} = {\cal R}^{(0)}{}^S{}_{MNP}\, g^{PQ} h_{QS}$ thanks to symmetries of the Riemann tensor. This expansion can now be used in the first order Einstein equation \eqref{Einstein1}. In the following, we assume that the background (i.e.~zeroth order) equation \eqref{Einstein0} is satisfied. Using it to replace ${\cal R}^{(0)}_{MP}$, the first order Einstein equation becomes
\bea
- \frac{1}{2} \square^{(0)}_D\ h_{MN} + {\cal R}^{(0)}{}^S{}_{MNP}\, g^{PQ} h_{QS} + \nabla^{(0)}_{(M} {\cal G}_{N)}  = 0 \ .\label{final1}
\eea

Finally, we impose the standard de Donder gauge in $D$ dimensions
\beq
{\cal G}_{N} = 0 \ ,\label{Dondergauge}
\eeq
where ${\cal G}_{N}$ is the de Donder operator defined in~\eqref{Donder}. This simplifies the first order Einstein equation \eqref{final1} to
\beq
- \frac{1}{2} \square^{(0)}_D\ h_{MN} + {\cal R}^{(0)}{}^S{}_{MNP}\, g^{PQ} h_{QS}  = 0 \ . \label{finalgen}
\eeq
Note that the $D$-dimensional de Donder gauge \eqref{Dondergauge} can always be reached locally whenever the Klein--Gordon equation with a source can be solved, because varying~\eqref{Donder} yields $\square^{(0)}_D \xi_M$ (up to terms proportional to the cosmological constant).\footnote{The existence of local solutions depends on the symbol of the operator, that is, on the term with the highest number of derivatives in the equation of motion. A study of solutions to the Klein--Gordon equation on curved space-times can be found e.g.~in~\cite{BosonsFermionsCurved}.} One may wonder whether there could be (unusual) global obstructions to imposing the de Donder gauge in the case where the extra dimensions are compact. Such potential global issues will however be ignored, since the above equations of motion are local, as will be, essentially, the subsequent analysis.

\subsection{Splitting dimensions and equations}\label{sec:split}

The dimensions are now split into $D=4+N$, i.e.~$4$ ``external'' dimensions corresponding to our extended space-time, and $N$ extra space dimensions. The latter are gathered as a manifold $\mmm$ and are dubbed ``internal'', even though we do not restrict for now to compact extra dimensions. The metric is decomposed accordingly. To that end, we consider for the background the most general metric that allows for the four-dimensional space-time to be maximally symmetric: this can be viewed as a physical requirement, as the four-dimensional space-time is then homogeneous and isotopic, and preserves Lorentz invariance. This standard metric \cite{vanNieuwenhuizen:1984cgg, VanNieuwenhuizen:1985be} corresponds to a warped product of a four-dimensional space-time of coordinates $x^{\mu}$ with $\mmm$ of coordinates $y^m$, meaning
\beq
{\rm Background:}\ \ \d s^2 = e^{2A(y)} \tilde{g}_{\mu \nu} (x) \d x^{\mu} \d x^{\nu} + g_{mn} (y) \d y^m \d y^n \ , \label{backgroundgen}
\eeq
where $e^{2A}$ is the warp factor and $g_{\mu\nu}= e^{2A} \tilde{g}_{\mu \nu}$. On the contrary, the fluctuation $h_{MN}$ does not need to be Lorentz invariant, so it is decomposed into $h_{\mu \nu}$, the ``vectors'' $h_{\mu m}$ and the ``scalars'' $h_{mn}$; its coordinate dependence is also generic. We introduce accordingly the traces $h_4=h_{\mu\nu} g^{\nu\mu}$, $\tilde{h}_4=h_{\mu\nu} \tilde{g}^{\nu\mu}$ and $h_N=h_{mn}g^{nm}$.

We now implement this information about the background and the dimensions in the wave equation \eqref{finalgen}. We first decompose the $D$-dimensional background connection coefficients: the only non-zero ones are
\begin{equation}
\begin{aligned}
& \Gamma^{M=\mu}_{NP=\nu\pi} = \tilde{\Gamma}^{\mu}_{\nu\pi} \ , \ \Gamma^{M=m}_{NP=np} = \Gamma^{m}_{np} \ ,\\
& \Gamma^{M=\mu}_{NP=\nu p} = \Gamma^{M=\mu}_{NP= p \nu} = \frac{1}{2} \delta^{\mu}_{\nu} e^{-2A} \del_p e^{2A} \ , \ \Gamma^{M=m}_{NP=\nu \rho} = -\frac{1}{2} \tilde{g}_{\nu \rho} g^{mn} \del_n e^{2A} \ ,
\end{aligned}
\end{equation}
where $\tilde{\Gamma}^{\mu}_{\nu\pi}$ is the four-dimensional coefficient built from the unwarped metric $\tilde{g}_{\mu\nu}$ and $\Gamma^{m}_{np}$ is the internal one built from $g_{mn}$. We then compute the $D$-dimensional background covariant derivatives acting on $h$,
\beal
\nabla_{Q=\rho}\, h_{MN=\mu\nu} &= \tilde{\nabla}_{\rho} h_{\mu\nu} + \tilde{g}_{\rho(\mu} h_{\nu) m} g^{mn} \del_n e^{2A} \ , \label{nabla4d}\\
\nabla_{Q=\rho}\, h_{MN=\mu n} &= \tilde{\nabla}_{\rho} h_{\mu n} + \frac{1}{2} \tilde{g}_{\rho \mu} h_{mn} g^{mp} \del_p e^{2A} - \frac{1}{2} h_{\mu\rho} e^{-2A} \del_n e^{2A} \ ,\\
\nabla_{Q=\rho}\, h_{MN= mn} &= \del_{\rho} h_{mn} - e^{-2A} h_{\rho (m} \del_{n)} e^{2A} \ ,
\eeal
where $\tilde{\nabla}_{\rho}$ is the purely four-dimensional, background-covariant derivative built from $\tilde{\Gamma}^{\mu}_{\nu\pi}$, and
\beal
\nabla_{Q=q}\, h_{MN=\mu\nu} &= \del_q h_{\mu\nu} - h_{\mu \nu} e^{-2A} \del_q e^{2A} \ ,\\
\nabla_{Q=q}\, h_{MN=m \nu} &= \nabla_{q} h_{m \nu} - \frac{1}{2} h_{m \nu} e^{-2A}\del_q e^{2A}\ ,\\
\nabla_{Q=q}\, h_{MN= mn} &= \nabla_{q} h_{mn} \ ,
\eeal
where $\nabla_{q}$ is the purely internal, background-covariant derivative built from $\Gamma^{m}_{np}$. Let us now compute the components of $\square_D^{(0)} h_{MN}$, where we recall that $\square_D^{(0)} = g^{PQ} \nabla_P\nabla_Q $ is built from the background metric. With $\tilde{\square}_4 = \tilde{g}^{\mu\nu} \tilde{\nabla}_{\mu} \tilde{\nabla}_{\nu}$, and the internal Laplacian $\Delta_{\mmm}=g^{pq} \nabla_{p} \nabla_{q}$, we obtain
\beal
\square_D^{(0)} h_{MN=\mu\nu} &= \ e^{-2A} \tilde{\square}_4 h_{\mu\nu} + \Delta_{\mmm} h_{\mu\nu} - h_{\mu\nu} \Delta_{\mmm} \ln e^{2A} -\frac{3}{2} e^{-4A} h_{\mu\nu}  g^{mn} \del_m e^{2A} \del_n e^{2A} \nn\\
& +2 e^{-2A} \tilde{\nabla}_{(\mu} h_{\nu) m} g^{mn} \del_n e^{2A} +\frac{1}{2} e^{-2A} \tilde{g}_{\mu\nu} h_{mn} g^{mr} g^{np} \del_r e^{2A} \del_p e^{2A} \ , \label{boxh4d}\\
\square_D^{(0)} h_{MN=\mu n} &= \ e^{-2A} \tilde{\square}_4 h_{\mu n} + \Delta_{\mmm} h_{\mu n} + e^{-2A} g^{pq} \nabla_p h_{\mu n} \del_q e^{2A}\\
&-\frac{3}{2} e^{-4A} h_{\mu m} g^{mp} \del_p e^{2A} \del_n e^{2A} - e^{-4A} h_{\mu n} g^{pq} \del_p e^{2A} \del_q e^{2A} - \frac{1}{2} h_{\mu n} \Delta_{\mmm} \ln e^{2A} \nn\\
&- e^{-4A} \tilde{g}^{\pi \rho} \tilde{\nabla}_{\pi} h_{\mu \rho} \del_n e^{2A} + e^{-2A} g^{pq} \del_{\mu} h_{np} \del_q e^{2A} \ ,\nn\\
\square_D^{(0)} h_{MN=mn} &=\ e^{-2A} \tilde{\square}_4 h_{mn} + \Delta_{\mmm} h_{mn} + 2 e^{-2A} g^{pq} \del_p e^{2A} \nabla_q h_{mn} - 2 e^{-4A} g^{pq} \del_p e^{2A} h_{q(m} \del_{n)} e^{2A} \nn\\
&- 2 e^{-4A} \tilde{g}^{\pi \rho} \tilde{\nabla}_{\pi} h_{\rho (m} \del_{n)} e^{2A} + \frac{e^{-4A}}{2}  h_4 \del_m e^{2A} \del_n e^{2A} \ .
\eeal
The components of the background Riemann tensor are also computed: they are given by
\begin{subequations}
	\begin{align}
{\cal R}^{(0)}{}^P{}_{MN=\mu\nu \ S} &= \delta^{\sigma}_{S}\delta^{P}_{\pi} \left(\tilde{{\cal R}}^{\pi}{}_{\mu\nu\sigma} + \tfrac{1}{2} e^{-2A} \delta^{\pi}_{[\sigma} \tilde{g}_{\nu]\mu} g^{pq} \del_p e^{2A} \del_q e^{2A} \right)\label{Riemann4d}\\
&+ \delta^{s}_{S}\delta^{P}_{n} \tfrac{1}{2} \tilde{g}_{\mu \nu} g^{np} \left(  \nabla_s \del_p e^{2A} -  \tfrac{1}{2} e^{-2A} \del_p e^{2A} \del_s e^{2A}  \right)\,, \nn\\
{\cal R}^{(0)}{}^P{}_{MN=\mu m \ S} &= \delta^{\sigma}_{S}\delta^{P}_{n} \tfrac{1}{2} \tilde{g}_{\sigma \mu} g^{np} \left( -\nabla_m \del_p e^{2A} + \tfrac{1}{2} e^{-2A} \del_p e^{2A} \del_m e^{2A}  \right)\,,\label{Riemannoffdiag}\\
{\cal R}^{(0)}{}^P{}_{MN=mn \ S} &= \delta^{s}_{S}\delta^{P}_{p} {\cal R}^{p}{}_{mns} \label{Riemannintern}\\
&+ \delta^{\sigma}_{S}\delta^{P}_{\pi} \tfrac{1}{2} \delta^{\pi}_{\sigma} \left( \nabla_n(e^{-2A} \del_m e^{2A} ) + \tfrac{1}{2} e^{-4A} \del_m e^{2A} \del_n e^{2A} \right) \nn \ .
	\end{align}
\end{subequations}
The components ${}_{MN}$ of the wave equation \eqref{finalgen} thus read
\begin{subequations}\label{eqgen}
	\begin{align}
{}_{\mu\nu}:\ & e^{-2A} \tilde{\square}_4 h_{\mu\nu} + \Delta_{\mmm} h_{\mu\nu} - h_{\mu\nu} \Delta_{\mmm} \ln e^{2A} \label{eqgen4d}\\
& -2 \tilde{{\cal R}}^{\pi}{}_{\mu\nu\sigma} g^{\sigma \rho} h_{\rho \pi} - \frac{1}{2} e^{-2A} g^{pq} \del_p e^{2A} \del_q e^{2A} \left( \tilde{g}_{\nu\mu} h_4 +2 h_{\nu\mu} e^{-2A} \right) \nn\\
& +2 e^{-2A} \tilde{\nabla}_{(\mu} h_{\nu) m} g^{mn} \del_n e^{2A} - \tilde{g}_{\mu\nu} h_{mn} g^{mr} g^{np} \left( \nabla_r \del_p e^{2A} - e^{-2A}\del_r e^{2A} \del_p e^{2A} \right) = 0 \,,\nn\\
{}_{\mu n}:\ & e^{-2A} \tilde{\square}_4 h_{\mu n} + \Delta_{\mmm} h_{\mu n} + e^{-2A} g^{pq} \nabla_p h_{\mu n} \del_q e^{2A} + e^{-2A} h_{\mu m} g^{mp} \nabla_n\del_p e^{2A} \label{eqgenoffdiag}\\
& -2 e^{-4A} h_{\mu m} g^{mp} \del_p e^{2A} \del_n e^{2A} - e^{-4A} h_{\mu n} g^{pq} \del_p e^{2A} \del_q e^{2A} - \frac{1}{2} h_{\mu n} \Delta_{\mmm} \ln e^{2A} \nn\\
& - e^{-4A} \tilde{g}^{\pi \rho} \tilde{\nabla}_{\pi} h_{\mu \rho} \del_n e^{2A} + e^{-2A} g^{pq} \del_{\mu} h_{np} \del_q e^{2A} = 0 \,,\nn\\
{}_{m n}:\ & e^{-2A} \tilde{\square}_4 h_{mn} + \Delta_{\mmm} h_{mn} + 2 e^{-2A} g^{pq} \del_p e^{2A} \nabla_q h_{mn} - 2 e^{-4A} g^{pq} \del_p e^{2A} h_{q(m} \del_{n)} e^{2A} \label{eqgenintern}\\
& -2 {\cal R}^s{}_{mnp} g^{pq} h_{qs} - 2 e^{-4A} \tilde{g}^{\pi \rho} \tilde{\nabla}_{\pi} h_{\rho (m} \del_{n)} e^{2A} - h_4 \nabla_n(e^{-2A} \del_m e^{2A} )= 0  \,.\nn
	\end{align}
\end{subequations}
These equations have been obtained in the $D$-dimensional de Donder gauge \eqref{Dondergauge}. This last condition also splits as follows:
\begin{subequations}
\label{DD12}
	\begin{align}
e^{-2A} \tilde{g}^{\pi\rho} \tilde{\nabla}_{\pi} h_{\rho \nu} - \frac{1}{2} e^{-2A} \tilde{\nabla}_{\nu} \tilde{h}_4 - \frac{1}{2} \nabla_{\nu} h_N + g^{pq} \nabla_p h_{q\nu} + 2 h_{p\nu} g^{pq} e^{-2A} \del_q e^{2A} & = 0 \ , \label{DD1}\\
g^{pq} \nabla_p h_{qr} - \frac{1}{2} \nabla_r h_N - \frac{1}{2} e^{-2A} \nabla_r \tilde{h}_4 + g^{\pi\rho} \tilde{\nabla}_{\pi} h_{\rho r} + 2 h_{mr} g^{mp} e^{-2A} \del_p e^{2A} & = 0 \ . \label{DD2}
	\end{align}
\end{subequations}

In view of the above wave equations, let us make a first comment on the idea of getting an amplitude damping. An exponential decrease of the wave amplitude along its propagation would be due to a dissipative term in the four-dimensional wave equation \eqref{eqgen4d}, i.e.~a term of the form $\tilde{\nabla}_{\mu} h_{\pi\rho}$ times an internal quantity. There is no such term in \eqref{eqgen4d};\footnote{Such a term is present, through a divergence, in the off-diagonal equation \eqref{eqgenoffdiag}. $h_{\mu n}$ being also present in \eqref{eqgen4d}, one may wonder whether this component mixing could not eventually generate a dissipative term in \eqref{eqgen4d}. The number of derivatives is however not the right one for this to happen.} in fact, with diffeomorphism invariance, linearization and the form of the background, one can show that such a term cannot be present. In other words, extra dimensions do not lead to a four-dimensional amplitude damping.

Before analysing further these equations in the next section, let us compute the cosmological constants using the above results. First, with the cosmological constant $\Lambda_4 = \tfrac{1}{4} \tilde{{\cal R}}_4$, the Riemann tensor is fixed as follows by considering our four-dimensional space-time to be maximally symmetric:
\beq
\tilde{{\cal R}}^{\pi}{}_{\mu\nu\sigma} = \frac{\Lambda_4}{3} \left( \delta^{\pi}_{\nu} \tilde{g}_{\mu\sigma} -  \delta^{\pi}_{\sigma} \tilde{g}_{\mu\nu}  \right) \ ,
\eeq
giving for \eqref{eqgen4d}
\beq
\tilde{{\cal R}}^{\pi}{}_{\mu\nu\sigma} g^{\sigma \rho} h_{\rho \pi} = e^{-2A}\frac{\Lambda_4}{3} \left( h_{\mu \nu} -  \tilde{h}_4\, \tilde{g}_{\mu\nu}  \right) \ . \label{bla}
\eeq
Second, thanks to the components of the background Riemann tensor \eqref{Riemann4d}, \eqref{Riemannoffdiag} and \eqref{Riemannintern}, we compute the Ricci tensor, and scalar ${\cal R}_D^{(0)}= - g^{MN} {\cal R}^{(0)}{}^P{}_{MNP}$. The $D$-dimensional cosmological constant is then given by
\beq
\frac{2D}{D-2}\, \Lambda_D = {\cal R}_D^{(0)} = e^{-2A} \tilde{{\cal R}}_4 + {\cal R}_{\mmm} - e^{-4A} (\del e^{2A})^2 - 4 e^{-2A} \Delta_{\mmm} e^{2A} \ ,\label{LD}
\eeq
where $( \del e^{2A} )^2 = g^{pq} \del_p e^{2A} \del_q e^{2A}$ and ${\cal R}_{\mmm}$ is the purely internal background Ricci scalar built from $g_{mn}$. Computing in addition
\beq
g^{\mu\nu} {\cal R}^{(0)}_{MN=\mu\nu} = e^{-2A} \tilde{{\cal R}}_4 -2 e^{-4A} (\del e^{2A})^2 - 2 e^{-2A} \Delta_{\mmm} e^{2A} \ ,
\eeq
we obtain the four-dimensional trace of equation \eqref{Einstein0}. This eventually gives
\beq
4 \Lambda_4 = \tilde{{\cal R}}_4 = \frac{4}{D-4} e^{2A} {\cal R}_{\mmm} + 2 \frac{D-2}{D-4} e^{-2A} (\del e^{2A})^2 + 2 \frac{D-8}{D-4} \Delta_{\mmm} e^{2A} \ . \label{L4}
\eeq
The $D$-dimensional and $4$-dimensional cosmological constants can then be compared: they are given by different expressions, namely different combinations of internal quantities, and further differ by an overall factor of $e^{2A}$. Therefore, the two cosmological constants can have different values: in particular, a small $e^{2A}$ would create a hierarchy between the two, i.e.~$\Lambda_4 << \Lambda_D$. This is the reason why the flat-space-time approximation usually made when studying gravitational-wave propagation cannot be justified in this $D$-dimensional context. Indeed, in four dimensions and for non-primordial gravitational waves, the typical length scale of variation of the perturbation is much smaller than that of the background, so that $\del^2 h_{\mu\nu} >> \Lambda_4 h_{\mu\nu}$ and the equation of motion reduces to the one on Minkowski. In the present $D$-dimensional setup, this reasoning breaks down if $\Lambda_4 << \Lambda_D$,\footnote{Note also that $h_{MN}$ varies also along the extra dimensions, leading a priori to different length scales, so that the comparison of scales is not as simple as in four dimensions.} so that we are to deal with the complete $D$-dimensional equation of motion at first order, considering a generic background as in \eqref{gDdecompo} and \eqref{backgroundgen}.

\section{Equation analysis and effects in four dimensions}\label{sec:constantwarp}

We have derived the wave equations \eqref{eqgen} describing the propagation of gravitational waves on the general background \eqref{backgroundgen}, which is the warped product of a four-dimensional space-time with $N$ extra dimensions. In the present section, we analyse them and determine the impact of extra dimensions on the four-dimensional wave.

For the sake of simplicity, the warp factor is taken constant from now on, and we relegate the study of the non-constant case to Appendix \ref{ap:nonconstantwarp}. Note that this corresponds to the common smearing approximation in supergravity. We then set $A=0$, since a constant $A$ can always be recovered by rescaling the four-dimensional metric. Accordingly, we drop the tilde notation introduced in \eqref{backgroundgen} and below, i.e.~we identify for instance $\tilde{g}_{\mu\nu}$ with $g_{\mu\nu}$, etc. Equations \eqref{eqgen} then boil down to
\begin{subequations}\label{constantwarp}
	\begin{align}
{}_{\mu\nu}:\hspace*{9,5pt} \square_4 h_{\mu\nu} + \Delta_{\mmm} h_{\mu\nu} &= 2 {\cal R}^{\pi}{}_{\mu\nu\sigma} g^{\sigma \rho} h_{\rho \pi}\,, \label{constantwarp4d} \\
{}_{\mu n}:\hspace*{8pt} \square_4 h_{\mu n} + \Delta_{\mmm} h_{\mu n} &= 0\,, \\
{}_{m n}:\;  \square_4 h_{mn} + \Delta_{\mmm} h_{mn} &= 2 {\cal R}^s{}_{mnp} g^{pq} h_{qs}\,.\label{constantwarp4dinternal}
	\end{align}
\end{subequations}
These equations were obtained in the $D$-dimensional de Donder gauge~\eqref{DD12}, which reads as follows when $A$ is set to zero:
\begin{subequations}
\label{TTDdimensions}
\begin{align}
g^{\pi\rho} \nabla_{\pi} h_{\rho \nu} + g^{pq} \nabla_p h_{q\nu} - \frac{1}{2} \nabla_\nu h_4 - \frac{1}{2} \nabla_\nu h_N & = 0\,,\label{TT2}\\
g^{\pi\rho} \nabla_{\pi} h_{\rho r} + g^{pq} \nabla_p h_{qr} - \frac{1}{2} \nabla_r h_4 - \frac{1}{2} \nabla_r h_N & = 0 \,.\label{TT3}
\end{align}
\end{subequations}
These gauge conditions together with equations \eqref{constantwarp} will be the starting point for the next two subsections. As a side remark, note that starting rather with the $D$-dimensional Transverse-Traceless gauge discussed in Appendix \ref{ap:TT}, one would obtain the same equations of motion \eqref{constantwarp}, but different gauge conditions. The results of the following two subsections would however remain the same.

Turning to the extra dimensions, we focus on the case of a compact $\mmm$ (without boundary). The internal Laplacian then admits a discrete orthonormal basis of eigenfunctions denoted $\{ \omega_{{\bf k}} (y) \}$ of discrete label ${\bf k}$, such that $\Delta_{\mmm}\, \omega_{{\bf k}} = - m_{\bf k}^2\, \omega_{{\bf k}}$ with a real $m_{\bf k}$. By convention, $m_{\bf 0}=0$. As an example, for $\mmm$ being a flat torus $T^N$, $\omega_{{\bf k}} (y) = e^{\i {\bf k} \cdot {\bf y}}$ and $\{{\bf k}\}$ is a set of $N$-dimensional real vectors isomorphic to $\mathbb{Z}^N$; another example can be found in~\cite{Andriot:2016rdd} with $\mmm$ being a nilmanifold. The general wave is then developed on this basis as
\beq
h_{MN} (x,y) = \sum_{{\bf k}} h_{MN}^{{\bf k}}(x)\, \omega_{{\bf k}} (y) \ , \label{decompogen}
\eeq
where $h_{MN}^{{\bf k}}$ are the Kaluza--Klein modes. Indeed, $h_{\mu\nu}$ is a scalar from the internal perspective, and each component of the internal tensors $h_{m\nu}$ and $h_{mn}$ can be viewed as a function too, and all of them can then be decomposed on this basis.

Before proceeding further with this mode expansion, we first indicate a few useful properties of the various modes, starting with the zero-mode. Harmonic functions $f$ on a compact manifold are constant, as can be viewed by integrating $f \Delta_{\mmm} f$. As a consequence, $\omega_{{\bf 0}}$ is constant and unique. More precisely, the zero-mode is always the constant part of $h_{MN}$ (with respect to $y^m$). Another manner of obtaining the zero-mode of a function is to integrate it over $\mmm$. Indeed, integrating an expansion over the basis $\{ \omega_{{\bf k}} \}$ only leaves the zero-mode because any $\omega_{{\bf k\neq 0}}$ integrates to zero; the latter can be seen by integrating to zero the total derivative $\Delta_{\mmm}\, \omega_{{\bf k}} = - m_{\bf k}^2\, \omega_{{\bf k}}$. As a consequence, any total derivative $\nabla_m X^m$ has a vanishing zero-mode since it integrates to zero (by the Gauss theorem). This property will be useful in the following.

More generally, to extract the ${\bf k}$-mode of a function or a scalar equation, one uses the orthonormality of the basis, namely multiplying an expansion by $\omega_{{\bf k}}^*$ and integrating over $\mmm$. For instance, using the expansion \eqref{decompogen} for $h_{\mu\nu}$ in \eqref{constantwarp4d} and the orthonormality, one gets for each mode
\beq
\label{eqk}
\square_4 h_{\mu\nu}^{{\bf k}} - m_{\bf k}^2\, h_{\mu\nu}^{{\bf k}} = 2 {\cal R}^{\pi}{}_{\mu\nu\sigma} g^{\sigma \rho} h_{\rho \pi}^{{\bf k}} \ .
\eeq
We will proceed analogously in the coming subsections. This mode decomposition leads to a splitting of the above equations of motion and gauge conditions into an infinite tower of equations and gauge conditions. In the next subsection we focus on the zero modes, whereas subsection~\ref{subsec:massive} will deal with the higher, Kaluza--Klein modes. The former will be understood as massless modes, while the latter will correspond to massive ones.

Finally, another background specification, to be made in the coming subsections, will be to consider the four-dimensional space-time to be Minkowski. This can be understood as a physical approximation, which suits the currently observed gravitational waves: in short, on the distances probed, the curvature of our universe is negligible; see also the discussion below \eqref{L4}. Considering Minkowski will in addition yield various technical simplifications. We now turn to the study of the various modes.

\subsection{Massless mode as a modified four-dimensional gravitational wave}\label{sec:massless}

Let us address the zero-modes, that is, we study the equations of motion~\eqref{constantwarp} and the gauge conditions~\eqref{TTDdimensions} for the $y^m$-independent contributions, following the procedure described above.

It is useful to analyse the gauge conditions first. Thanks to the previously discussed properties, the conditions~\eqref{TTDdimensions} are greatly simplified on the zero modes. From \eqref{TT2}, we obtain
\begin{equation}
g^{\pi\rho} \nabla_{\pi} h^{\bf 0}_{\rho \nu} - \frac{1}{2} \nabla_\nu h^{\bf 0}_4 = \frac{1}{2} \nabla_\nu h^{\bf 0}_N\,,\label{TT2zero}\\
\end{equation}
and we will come back to \eqref{TT3}. It is crucial to note that the four-dimensional zero-mode $h^{\bf 0}_{\mu \nu}$ fails to satisfy the four-dimensional de Donder gauge condition in \eqref{TT2zero}, because of the presence of $h^{\bf 0}_N \equiv (g^{mn}h_{mn})^{\bf 0}$, the zero-mode of the internal trace. In spite of the system of equations \eqref{constantwarp} being diagonal regarding the various $h_{MN}$ components, those actually mix and are related to each other through the above gauge conditions. This is the source of the effect which we unveil in this subsection, namely, the four-dimensional gravitational wave $h^{\bf 0}_{\mu \nu}$ is modified by the presence of $h^{\bf 0}_N$.

To identify the effect clearly we now approximate, as mentioned previously, the four-dimensional space-time to be Minkowski. We set $g_{\mu\nu} = \eta_{\mu\nu}$ so that, in particular, the right-hand side of equation~\eqref{constantwarp4d} vanishes. Combining the background equations \eqref{Einstein0}, \eqref{LD} and \eqref{L4}, one sees that the restriction to Minkowski forces the background internal geometry to be Ricci flat, i.e.~${\cal R}_{mn}=0$. Interestingly, this Ricci tensor appears when tracing equation \eqref{constantwarp4dinternal} by a contraction with $g^{mn}$: the resulting right-hand side then vanishes. Thus we can focus only on the wave $h^{\bf 0}_{\mu \nu}$ together with the internal trace $h^{\bf 0}_N$, since they decouple completely from the rest. The components $h^{\bf 0}_{\mu \nu}$ and $h^{\bf 0}_N$ can also be checked not to appear in \eqref{TT3}, which justifies why the study of this second gauge condition can be omitted. As the internal Laplacian term has no zero mode in equations \eqref{constantwarp}, we are left with the following system:
\begin{empheq}{align}
\square_4 h^{\bf 0}_{\mu\nu} &= 0\,, \label{constantwarp4d0mode} \\
\square_4 h^{\bf 0}_N &= 0\,,\label{constantwarp4dinternal0mode}
\end{empheq}
together with \eqref{TT2zero}. One recognises the usual equations of motion for a free massless graviton (as for a standard gravitational wave) along with a free massless scalar field on Minkowski space-time. However, as we will now detail, the gauge condition~\eqref{TT2zero} will crucially change the polarization properties of the gravitational wave $h^{\bf 0}_{\mu\nu}$.\\

Before proceeding any further, it is instructive to perform the counting of degrees of freedom and determine the residual gauge invariance for the zero modes. We will only discuss the system formed by $h^{\bf 0}_N $ together with $h^{\bf 0}_{\mu\nu}$. Firstly, it is easy to see that the former is a scalar field. Indeed, from the transformation rule~\eqref{gaugevariationgeneric} for the generic wave, one sees that the internal trace $h_N$ transforms according to $h_N \to h_N + 2 g^{mn}\nabla_m \xi_n$, which implies that the zero mode  $h^{\bf 0}_N$ does not transform: $\delta h^{\bf 0}_N = 0$ (as explained after \eqref{decompogen}, $\nabla_m (g^{mn}\xi_n)$ has a vanishing zero-mode). For $h^{\bf 0}_{\mu\nu}$, taking the zero-mode of the full transformation rule $\delta h_{\mu\nu} = 2\partial_{(\mu}\xi_{\nu)}$ shows that $h^{\bf 0}_{\mu\nu}$ transforms as a spin-2 field with infinitesimal gauge parameter $\xi^{\bf 0}_{\nu}$:
\begin{equation}
\label{gaugevariationzero4d}
	\delta h^{\bf 0}_{\mu\nu} = 2\partial_{(\mu}\xi^{\bf 0}_{\nu)}\,.
\end{equation}
However, since the gauge condition~\eqref{TT2zero} deviates from the usual de Donder gauge condition in four dimensions, one may ask whether the components $h^{\bf 0}_{\mu\nu}$, although they transform as a graviton and obey the standard equation~\eqref{constantwarp4d0mode}, really are to be regarded as a graviton. The key observation is that, since $h^{\bf 0}_N$ does not transform, the condition~\eqref{TT2zero} leaves one with the exact same residual gauge freedom as when imposing the usual de Donder gauge in four dimensions, that is, $\square_4 \xi^{\bf 0}_\mu = 0$. This is seen by taking a gauge variation of \eqref{TT2zero} following \eqref{gaugevariationzero4d}: the right-hand side of~\eqref{TT2zero} does not transform and yields zero, while the left-hand side yields $\square_4 \xi^{\bf 0}_\mu$. Note that the residual gauge condition $\square_4 \xi^{\bf 0}_\mu = 0$ ensures that the equation of motion~\eqref{constantwarp4d0mode} is gauge invariant, as it should.

We conclude that $h^{\bf 0}_{\mu\nu}$ obeying~\eqref{constantwarp4d0mode} is indeed a four-dimensional spin-$2$ field, which is however coupled to the scalar field $h^{\bf 0}_N $ in an unusual way, namely, via the condition~\eqref{TT2zero} inherited from the higher-dimensional de Donder gauge imposed in~\eqref{Dondergauge}. In particular it is then clear that the four-dimensional wave carries $2$ degrees of freedom. Indeed, the standard counting goes through despite the non-trivial right-hand side in~\eqref{TT2zero}: the symmetric tensor $h^{\bf 0}_{\mu\nu}$ originally has $10$ independent components, from which the condition~\eqref{TT2zero} subtracts $4$.\footnote{In the condition~\eqref{TT2zero} the right-hand side may be viewed as a source: the left-hand side, instead of being set to zero, is set to some non-zero, fixed quantity which does not transform. Setting this ``source'' to zero, which is consistent with equation~\eqref{constantwarp4dinternal0mode}, turns~\eqref{TT2zero} into the usual de Donder condition in four dimensions.} The residual gauge invariance $\square_4 \xi^{\bf 0}_\mu = 0$ can be further used to remove $4$ additional components, yielding the usual $2$ degrees of freedom of a massless graviton. In the following we will perform this reduction explicitly. \\

We now show precisely how the presence of $h^{\bf 0}_N $ in the condition~\eqref{TT2zero} affects the polarization properties of the gravitational wave $h^{\bf 0}_{\mu\nu}$. In order to do this we expand the four-dimensional wave as well as the scalar $h^{\bf 0}_N $ on a basis of real solutions to \eqref{constantwarp4d0mode} and \eqref{constantwarp4dinternal0mode}, namely plane-waves with a wave vector $k^{\rho}$ that is light-like
\begin{subequations}
\label{planewavedecomposition}
\begin{align}
	h^{\bf 0}_{\mu\nu} &= \int \d^4 k\ e^k_{\mu\nu}\, {\rm Re} \{\textrm{e}^{\i k_\rho x^\rho}\}\,,\\
	h^{\bf 0}_N  &= \int \d^4 k\ f_N^k\, {\rm Re} \{\textrm{e}^{\i k_\rho x^\rho}\}\,,
\end{align}
\end{subequations}
with a sum over $k_\rho k^\rho = 0$. The complex exponential is projected by ${\rm Re}$ on its real part. For simplicity we do not consider the imaginary part and its coefficient, which forms another set of independent basis elements: those would lead to a similar effect as the one we point-out. Plugging the above expansion into condition \eqref{TT2zero} (which amounts to a Fourier transform), we obtain for each $k^{\rho}$
\begin{equation}
e^k_{\mu\nu} k^\mu - \frac{1}{2} e_4^k k_\nu = \frac{1}{2} f_N^k k_\nu \,,
\end{equation}
where $e_4^k \equiv \eta^{\mu\nu}e^k_{\mu\nu}$. Since the wave vector is light-like, we now choose $k^\rho = (\omega / c,0,0,k)$ with the angular frequency $\omega= kc \geq 0$, which amounts to rotating our coordinate system so that the wave $h^{\bf 0}_{\mu\nu}$ propagates in the $x^3$ direction. Note that we only consider the left-traveling (or retarded) wave. For convenience let us set here $c = 1$, $\omega = 1$, and drop the label $k$; we will restore them later. We thus specify the third equation above for $\nu = 0,1,2,3$ with $k^\rho = (1,0,0,1)$, which yields
\begin{subequations}
\label{TTplanewave}
\begin{align}
	e_{00} + e_{03} + \tfrac{1}{2} e_4 &=  -\tfrac{1}{2} f_N\,,\\
	e_{01} + e_{13} &= 0\,,\\
	e_{02} + e_{23} &= 0\,,\\
	e_{33} + e_{03} - \tfrac{1}{2} e_4 &=  \tfrac{1}{2} f_N\,.
	\end{align}
\end{subequations}

We now fix the gauge further using the residual gauge invariance $\delta h^{\bf 0}_{\mu\nu} = 2\partial_{(\mu}\xi^{\bf 0}_{\nu)}$ with $\square_4 \xi^{\bf 0}_{\nu} = 0$. Thanks to the latter, we also expand the gauge parameter $\xi^{\bf 0}_{\nu}$ in real plane-waves with light-like wave vectors $k^\rho$, as $\xi^{\bf 0}_{\nu} = \int \d^4 k\,\chi^k_{\nu}\, {\rm Im} \{\textrm{e}^{\i k_\rho x^\rho}\}$, where ${\rm Im}$ projects on the imaginary part. The $h^{\bf 0}_{\mu\nu}$ transformation rule then reads $\delta e^k_{\mu\nu} = 2 k_{(\mu}\chi^k_{\nu)}$ for each $k^{\rho}$, where $\chi^k_\nu$ is arbitrary. We drop again the $k$ labels from now on. By inspection of this transformation rule one sees that $\chi_\nu$ can always be fixed so as to set $e_{0\nu} = 0$. The conditions~\eqref{TTplanewave} above then imply that $e_{3i} = 0$ for $i = 1,2,3$, as well as $e_4 = e_{11} + e_{22} = - f_N$. The polarization matrix $e_{\mu\nu}$ eventually reads
\begin{equation}
e_{\mu\nu} =
	\begin{pmatrix}
		0 & 0 & 0 & 0\\
		0 & e_{11} & e_{12} & 0\\
		0 & e_{12} & -e_{11} - f_N & 0\\
		0 & 0 & 0 & 0
	\end{pmatrix}\,,
\end{equation}
where the gauge is now completely fixed. Note that this result was mentioned in~\cite{Alesci:2005rb}, in the simplest setup where $\mathcal{M}$ is a circle. We now provide in the following further interpretation of this effect. \\

To start with, the plane-wave can be rewritten as
\begin{empheq}[innerbox=\fbox,left=\!\!\!\!]{align}
h^{\bf 0}_{ab}(t,x^3) =
	\begin{pmatrix}
		h^+ - \frac{1}{2} f_N & h^\times\\
		h^\times & -h^+ - \frac{1}{2} f_N
	\end{pmatrix}_{ab} \textrm{cos}(\omega (t - x^3 /c)) \equiv h^{\times}_{ab} + h^{+}_{ab} + h^{\ocircle}_{ab} \label{gusmatrix}
\end{empheq}
where $a, b = 1, 2$ and we have restored the speed of light $c$ and the angular frequency $\omega$. This is the modified gravitational wave in a gauge analogue to the Transverse-Traceless gauge. However since the field is obviously not traceless, we refer to the above gauge as the Transverse-Trace-fixed gauge, the trace being fixed and given by $-f_N$. As anticipated, this gravitational wave carries $2$ degrees of freedom, the two free constants $h^+$ and $h^\times$, while $f_N$ is a fixed, independent quantity. The latter gives rise to a so-called breathing mode $h^{\ocircle}_{ab}$, which is transverse and exists on top and independently of the two standard polarization modes $h^{\times}_{ab}$ and $h^{+}_{ab}$. Generating a breathing mode is an effect that has been noticed before in other contexts, for instance in alternative theories of gravity. A review on this topic can be found in \cite{Will:2014kxa}.\footnote{\label{foot:scalartensor}The only theories having a breathing mode as the only extra massless mode with respect to General Relativity seem to be scalar-tensor gravities~\cite{Will:2014kxa}. A detailed comparison of these models to our setup would be interesting.} In our setup the presence of this extra mode is a consequence of having extra compact dimensions in the universe.

The effect of the breathing mode $h^{\ocircle}_{ab}$ is most easily understood by looking at the stretching and shrinking of space in the transverse plane induced by the above gravitational wave, with $f_N \neq 0$. The standard textbook derivation remains formally the same, using the well-known equation for the geodesic deviation in the proper detector frame:
\begin{equation}
\label{geodesicequation}
\ddot{{\cal E}}_a = \frac{1}{2}\ddot{h}^{\bf 0}_{ab}{\cal E}^{b}\,,
\end{equation}
where dots denote time-derivatives and ${\cal E}^a= x^a_0 + \Delta x^a $ is the transverse-plane deviation of one test-point geodesic with respect to another (we refer to~\cite{Maggiore:1900zz} for details).\footnote{One may question the use of \eqref{geodesicequation} in our setting. Formally, one should rather consider the equation for the geodesic deviation in $D$ dimensions, and look at contributions to the four-dimensional $\ddot{{\cal E}_{\mu}}$. The vector $h_{\mu m}$ then allows for terms depending on the internal ${\cal E}^{m}$ in the right-hand side. However, as an internal distance, ${\cal E}^{m}$ is typically be much smaller than ${\cal E}^{\mu}$, so that its contributions can be neglected. Further, $\Delta x^{\mu}$ in ${\cal E}^{\mu}$ can be decomposed on the basis $\{ \omega_{{\bf k}} \}$, and one can project on the zero-mode: the resulting equation is \eqref{geodesicequation}.} From the above matrix \eqref{gusmatrix}, the equation \eqref{geodesicequation} tells us that the breathing mode yields the following deformation of distances in the transverse plane $x^3 = 0$:
\begin{subequations} \label{eqbox1}
	\begin{align}
		\Delta x^1 (t) &= -\tfrac{1}{4} f_N ~x^1_0 ~\textrm{cos} (\omega t)\,,\\
		\Delta x^2 (t) &= -\tfrac{1}{4} f_N ~x^2_0 ~\textrm{cos} (\omega t)\,.
	\end{align}
\end{subequations}
We further comment on the observability of this effect in Section \ref{sec:ccl}.

\subsection{Massive modes as high-frequency signals}
\label{subsec:massive}

Let us now turn to the four-dimensional higher modes $h^{{\bf k } \neq {\bf 0}}_{\mu\nu}$. We study the equations of motion \eqref{constantwarp} and the gauge conditions \eqref{TTDdimensions} for the higher Kaluza--Klein modes of the expansion~\eqref{decompogen}, and restrict again the external space-time to be Minkowski, $g_{\mu\nu} = \eta_{\mu\nu}$. In line with the analysis of the massless wave, where the gauge conditions are non-trivial, we will first comment on the conditions obeyed by the higher modes and show that they indeed correspond to standard massive fields. We will then discuss the interpretation of these modes in terms of gravitational waves.\\

The equation of motion for the four-dimensional modes reads, from \eqref{eqk},
\begin{equation}
\label{massivemodeequation}
	\square_4 h^{\bf k }_{\mu\nu} - m^2_{\bf k} h^{\bf k }_{\mu\nu} = 0\,.
\end{equation}
This is the equation for a standard, transverse and traceless graviton of mass $m_{\bf k}$ on Minkowski space-time, where we set the speed of light $c$ and the reduced Planck constant $\hbar$ to $1$. However, the gauge conditions \eqref{TTDdimensions}, when written in terms of $h^{\bf k }_{\mu\nu}$, do not immediately imply that $h^{\bf k }_{\mu\nu}$ is Transverse-Traceless. Rather, \eqref{TT2} reads
\begin{equation}
		\partial^\nu h^{\bf k }_{\mu\nu} - \tfrac{1}{2}\partial_\mu h^{\bf k }_4 = -(g^{mn} \nabla_m h_{n\mu})^{\bf k } + \tfrac{1}{2}\partial_\mu h^{\bf k }_N\,,\label{firstgaugecondition}
\end{equation}
and the second gauge condition~\eqref{TT3} is left out, anticipating that it will not play any role. Again note that $g^{mn}$ may generically depend on the internal coordinates $y^m$, so that $(\nabla_m g^{mn}h_{n\mu})^{\bf k }$ includes in general different modes $h^{\bf n}_{m\nu}$.

It is evident from \eqref{firstgaugecondition} that $h^{\bf k }_{\mu\nu}$ does not satisfy a priori a de Donder condition, even less so a Transverse-Traceless condition. This is to be expected, since in fact $h^{\bf k }_{\mu\nu}$ and the other components enjoy non-trivial gauge variations, namely
\beq\label{massivegaugeconditions}
\delta h^{\bf k }_{\mu\nu} = 2 \partial_{(\mu}\xi^{\bf k }_{\nu)}\,,
\eeq
obtained as the $\bf k$-mode projection of the generic gauge variation~\eqref{gaugevariationgeneric}, and analogously for the variations of $h^{\bf k }_{m\nu}$ and $h^{\bf k }_{mn}$. The crucial difference with the massless case studied in the previous subsection is that, in \eqref{firstgaugecondition}, the right-hand side now also transforms under a gauge variation \eqref{gaugevariationgeneric} of the fields, while in~\eqref{TT2zero} the right-hand side did not transform. Actually, the four-dimensional mode $h^{\bf k }_{\mu\nu}$ is massive but it is cast in a Stueckelberg-like formalism \cite{Stueckelberg:1900zz,Stueckelberg:1938zz,Delbourgo:1975aj}. In the latter, gauge invariance is introduced for a massive field, which in principle has no corresponding gauge parameter, and its trace and divergence are non-zero. It is well known in this setup that one can completely fix the gauge freedom back so as to obtain the usual formulation of a massive field, that is, a Transverse-Traceless field obeying equation~\eqref{massivemodeequation} and enjoying no gauge invariance. This is what we prove now.

The first step is to determine the residual gauge invariance allowed by the condition~\eqref{firstgaugecondition}. It is sufficient to consider only the latter, as we are interested in $h^{\bf k }_{\mu\nu}$, which will eventually decouple from the rest. The gauge variation of equation~\eqref{firstgaugecondition} reads
\begin{equation}
	\label{massiveresidual}
	\square_4 \xi^{\bf k }_{\mu} = m_{\bf k}^2 \xi^{\bf k }_{\mu}\,.
\end{equation}
Note that one can use~\eqref{massiveresidual} to check that the equation of motion~\eqref{massivemodeequation} is left invariant, as it should. Now, we are going to fix the above residual freedom completely, by imposing the Transverse-Traceless condition on $h^{\bf k }_{\mu\nu}$, that is,\footnote{One can check straightforwardly that the gauge~\eqref{TTconditionsmassive} is reachable by a residual gauge transformation with a parameter obeying~\eqref{massiveresidual}.}
\begin{equation}
\label{TTconditionsmassive}
		\partial^\nu h^{\bf k }_{\mu\nu} = 0\,,\qquad h^{\bf k }_4 = 0 \,.
\end{equation}
Taking the variation of these equations, it is easy to see that gauge parameters which preserve the above conditions need to satisfy
\begin{equation}
		\square_4 \xi^{\bf k }_{\mu} + \partial_\mu \partial^\nu \xi^{\bf k }_{\nu} = 0\,,\qquad \partial^\nu \xi^{\bf k }_{\nu} = 0 \,.
\end{equation}
Combining these conditions with one another obviously yields $\square_4 \xi^{\bf k }_{\nu} = 0$. This in turn makes the condition~\eqref{massiveresidual} boil down to $m_{\bf k}^2 \xi^{\bf k }_{\mu} = 0$, which fixes $\xi^{\bf k }_{\nu}$ to vanish. It is then clear that the transverse and traceless conditions~\eqref{TTconditionsmassive} completely fix the four-dimensional gauge freedom, so that the massive field $h^{\bf k }_{\mu\nu}$ no longer enjoys a gauge transformation. Thereby we recover the familiar formulation of a massive spin-$2$ field, obeying the conditions~\eqref{TTconditionsmassive} and satisfying the equation of motion~\eqref{massivemodeequation}.\\

In the gauge \eqref{TTconditionsmassive} instead of \eqref{firstgaugecondition}, we recover a familiar, transverse and traceless massive graviton $h^{\bf k }_{\mu\nu}$ obeying the equation of motion~\eqref{massivemodeequation} and fully decoupled from the rest. We now write $h^{\bf k }_{\mu\nu}$ as the general solution to the massive Klein--Gordon equation \eqref{massivemodeequation}, i.e.~given in terms of plane-waves as
\begin{equation}
\label{fourriertransformmassive}
	h^{\bf k}_{\mu\nu} = \int \d^4 p_{\bf k}\, e^{p_{\bf k}}_{\mu\nu}\, {\rm Re} \{\textrm{e}^{\i p_{{\bf k}\rho} x^\rho}\}\,,
\end{equation}
where again we consider for simplicity only the real part, and the sum is over wave vectors $p_{\bf k}^\rho = (\omega_{\bf k}, \vec{p}_{\bf k})$ satisfying the massive dispersion relation
\begin{equation}
\label{dispersionmassive}
\omega_{\bf k}^2 = m^2_{\bf k} + \vec{p}^{\,2}_{\bf k}\,,	
\end{equation}
where $\omega_{\bf k} > 0 $ is the angular frequency of the wave. We recall that we have set $c=\hbar=1$.

Along the lines of the previous subsection, one can now determine the polarization matrix $e_{\mu\nu}$ for each $p_{\bf k}$. First let us note that the above dispersion relation allows us to choose $p_{\bf k}^\rho = (\omega_{\bf k}, 0,0,0)$. This amounts to picking a reference frame traveling along with the field, so that it appears static. With this choice of $p_{\bf k}^\rho$, plugging the expansion~\eqref{fourriertransformmassive} into the conditions~\eqref{TTconditionsmassive} implies $e_{0\nu} = 0$ and $\sum_{i} e_{ii} = 0$, so that the polarization matrix can be written down as
\begin{equation}
 	e_{\mu\nu} =
	\begin{pmatrix}
		0 & 0 & 0 & 0\\
		0 & e_{11} & e_{12} & e_{13}\\
		0 & e_{12} & -e_{11}-e_{33} & e_{23}\\
		0 & e_{13} & e_{23} & e_{33}
	\end{pmatrix}\,,
\end{equation}
which has $5$ independent components, as it should. Together they account for the massive wave $h^{{\bf k} \neq {\bf 0}}_{\mu\nu}$, which in the chosen reference frame is rewritten, for each~$\omega_{\bf k}$, as
\begin{empheq}[innerbox=\fbox, left=\!\!\!\!]{align} \label{eqbox2}
h^{{\bf k} \neq {\bf 0}}_{ij}(t) =
	\begin{pmatrix}
		h^+ -\tfrac{1}{2} h^{l,\ocircle} & h^\times & h^l_1\\
		 h^\times & -h^+ -\tfrac{1}{2} h^{l,\ocircle} & h^l_2\\
		 h^l_1 & h^l_2 & h^{l,\ocircle}
	\end{pmatrix}_{ij} \textrm{cos}(m_{\bf k}\,c^2\, t)
\end{empheq}
where the angular frequency is given by $m_{\bf k} c^2$, by virtue of~\eqref{dispersionmassive} with $c$ restored.  In the above, $h^+$ and $h^\times$ are two purely transverse modes, while $h^l_1$ and $h^l_2$ are two purely longitudinal ones (see \cite{Will:2014kxa} for the definition of the polarization modes). The fifth component, $h^{l,\ocircle}$, is mixed; in particular, in the plane transverse to $x^3$ it gives rise to a breathing mode. All these massive modes are part of the original four-dimensional wave $h_{\mu\nu}(x^\mu,y^m)$, and therefore add up to the (modified) gravitational wave $h^{\bf 0}_{\mu\nu}$ studied in the previous subsection. We further comment on the characteristics and observability of these extra signals in Section \ref{sec:ccl}; in particular, we will see that $m_{\bf k}$ is expected to dominate over $|\vec{p}_{\bf k}|$ in \eqref{dispersionmassive}, leading to signals of high frequency.

\section{Summary and discussion}\label{sec:ccl}

In this paper we have addressed the following question from a comprehensive perspective: \emph{can extra dimensions induce unusual effects on gravitational waves in four dimensions, and if so how are they characterized?} In this section we present our results and discuss their observability.

\subsection*{Results and comments}\addcontentsline{toc}{subsubsection}{Results and comments}

Starting from General Relativity with a cosmological constant in~$D=4+N$ dimensions on the most generic, warped space-time geometry satisfying the background equations of motion, we have derived the wave equations~\eqref{eqgen} for all components of $h_{MN}(x^\rho, y^m)$ from the $D$-dimensional linearized wave equation \eqref{finalgen}. We have done so by imposing the de Donder gauge condition in dimension~$D$ and then performing the reduction; we comment on different gauge choices here after. From the general equations \eqref{eqgen}, a first result is that $D$-dimensional General Relativity does not allow for an amplitude damping of the four-dimensional gravitational wave, even with extra dimensions, contrary to the ``extra-dimensional leakage'' observed in \cite{Deffayet:2007kf} that we attribute to modified gravity.

In the case of a background with a constant warp factor, approximating the four-dimensional geometry by Minkowski space-time (yielding a Ricci flat internal manifold), we have focused on the four-dimensional gravitational wave $h_{\mu\nu}$. This wave has a massless part given by its zero-mode $h^{\bf 0}_{\mu\nu}$ and a massive piece given by its non-zero modes $h^{{\bf k} \neq {\bf 0}}_{\mu\nu}$, and it differs from usual gravitational waves obtained from four-dimensional General Relativity in two ways:
\begin{enumerate}
	\item The massless wave $h^{\bf 0}_{\mu\nu}$ generically has an extra polarization mode on top of the ``plus'' and ``cross'' modes of General Relativity, namely a so-called breathing mode~\eqref{gusmatrix}, whose amplitude is determined by a specific scalar combination of the other components.
	\item The massive tower $h^{{\bf k} \neq {\bf 0}}_{\mu\nu}$ of extra waves \eqref{eqbox2} have all six polarization modes turned on, only five of them being independent, and they add up to the massless wave with a discrete spectrum of high frequencies.
\end{enumerate}
We explicitly show the zero-mode $h^{\bf 0}_{\mu\nu}$ to be a free massless spin-$2$ wave, while the Kaluza--Klein higher modes $h^{{\bf k} \neq {\bf 0}}_{\mu\nu}$ satisfy a standard massive equation of motion. The massless wave being free, let us emphasize that the extra scalar $h^{\bf 0}_N=(g^{mn}h_{mn})^{\bf 0}$ responsible for the massless breathing mode is absent from its equation of motion. Rather, it appears in the gauge conditions, which we analyze in great detail for both the massless and the massive modes. The massless wave is found to propagate only two degrees of freedom, in spite of the extra scalar which modifies its four-dimensional gauge conditions. This extra scalar does not transform under a gauge transformation, and therefore cannot be removed. The massive waves $h^{{\bf k} \neq {\bf 0}}_{\mu\nu}$ propagate five independent degrees of freedom each, but have all polarization modes turned on, with frequencies satisfying a massive dispersion relation.

Another important comment is that our results have been obtained by imposing the de Donder gauge in $D$ dimensions. However, as can be checked from the expressions of Appendix~\ref{ap:TT}, the above two results would remain unaltered if one starts with the $D$-dimensional Transverse-Traceless gauge instead. One could also proceed without imposing any gauge choice in $D$ dimensions. This would lead, a priori, to more complicated equations of motion for the zero-modes even in the Minkowskian, constant-warp factor case. Then imposing the four-dimensional de Donder or Transverse-Traceless gauge would yield couplings among the fields in the equations of motion, and in particular between $h^{\bf 0}_{\mu\nu}$ and $h^{\bf 0}_N$. It would be interesting to check that this leads to the same effect. In addition, this would allow to compare in detail our setup to four-dimensional scalar-tensor gravities, as motivated in footnote \ref{foot:scalartensor}.

Finally, more involved cases have been studied, namely considering a non-constant warp factor in Appendix \ref{ap:nonconstantwarp}, or a more general $D$-dimensional Lagrangian going beyond General Relativity in Appendix \ref{ap:gen}. In both cases, the complete equations of motion have been derived and discussed. For a non-constant warp factor, an interesting observation is that the four-dimensional Transverse-Traceless gauge conditions emerge naturally from the extra equations of motion, if one sets to zero the extra (vector and scalar) wave components. The resulting four-dimensional equation reproduces that obtained in previous works, captured by our general framework, including for instance those on a Randall--Sundrum background \cite{Randall:1999ee}. This suggests interesting generalisations beyond those works. We refer to the appendices for more details. We now discuss to what extent the two above effects can be observed.

\subsection*{Observability of the effects}\addcontentsline{toc}{subsubsection}{Observability of the effects}

We now focus in more detail on the physics associated with the two effects mentioned above. We explain why they could not have been detected so far, and discuss to what extent they could be observed in the future.\\

As reported, the first effect consists in having a breathing mode in the massless four-dimensional gravitational wave~\eqref{gusmatrix}. This is one of the six possible polarization modes, and one of the four that General Relativity does not predict, and is thus symptomatic of new physics \cite{Will:2014kxa}. The breathing mode deforms the space in a specific manner described by \eqref{eqbox1}, giving a distinct signature. To observe it, one needs to disentangle it from the other two, standard, transverse modes, which requires at least three, differently oriented detectors \cite{Will:2014kxa}. Currently, only the LIGO detector is active. In addition, its two sites have almost aligned arms, which maximizes its sensitivity but allows to detect only one polarization mode (see however \cite{Callister:2017ocg} on detecting polarization modes of the stochastic gravitational-wave background). More detectors should be available in a near future.

Observing this breathing mode would also require its amplitude to be high enough. This amplitude is given in \eqref{gusmatrix} for each plane wave by $f_N$, a quantity related to the four-dimensional Fourier coefficient of $h_N^{{\bf 0}}= (g^{mn}h_{mn})^{{\bf 0}}$. The latter is the zero-mode of the trace along the extra dimensions, meaning the constant part or mean value of this trace with respect to the extra coordinates. Then, for the breathing mode amplitude to be non-zero, one first needs some non-zero $h_{mn}$ to be emitted. This depends on the physics of the source in the extra dimensions, on which we have no control here. Assuming some non-zero $h_{mn}$, one may wonder about taking the trace, since one is used in four-dimensional General Relativity to get traceless waves. The discussion of Section \ref{sec:massless} makes it however clear that in our setup, neither the above internal trace $h_N$ nor the four-dimensional one need to vanish; in particular we showed that $h_N^{{\bf 0}}$ cannot be gauged away. Finally, one may wonder about the consequences of considering the zero-mode. If the internal metric $g_{mn}$ is constant as for a flat torus, one considers the zero-mode of $h_{mn}$ itself in the trace $h_N^{{\bf 0}}= \delta^{mn}h^{{\bf 0}}_{mn}$. Whether the latter vanishes depends again on unknown physics of the source, but we may still draw an analogy with known four-dimensional emissions, e.g.~by currently observed black hole mergers. In standard models, those produce a zero-mode of the four-dimensional wave. The latter is sometimes referred to as the gravitational wave memory effect \cite{Favata:2010zu}, which is less studied than the oscillatory part of the wave, perhaps because it is currently not detectable. This analogy plays in favor of an $h_{mn}^{{\bf 0}}$ being produced. Overall, we conclude that there is no physical reason for $h_N^{{\bf 0}}$ to be vanishing, generically leading to a non-zero breathing mode. An estimate of its amplitude remains however out of the scope of the present study.

Let us finally emphasize that $h_N^{{\bf 0}}$ is not sensitive to the size of the extra dimensions, because it is a zero-mode and it is thus independent of the extra coordinates. This is why it is here a massless mode. This is in contrast with the massive Kaluza--Klein modes, to be discussed. Including background fluxes or curvature on the internal manifold, one may generate a mass for $h_N^{{\bf 0}}$: this could be seen using equations of Appendix \ref{ap:gen}. With those equations, our analysis could be reproduced to determine whether the first effect occurs in that case. As for moduli in supergravity compactifications, such a mass should be lower than the first Kaluza--Klein mass, so it would not necessarily be related to the size of the extra dimensions. Therefore, the energy required to emit $h_N^{{\bf 0}}$, and generate a breathing mode, needs not be high.\\

We turn to the second effect, corresponding to extra four-dimensional waves, that have several unusual features depicted through \eqref{eqbox2}. First, they have all six polarization modes turned on (even if only five of them are independent). More importantly, all these polarization modes have in each plane wave the same angular frequency $\omega$ satisfying the massive dispersion relation \eqref{dispersionmassive}, where the label ${}_{{\bf k}}$ is dropped in the following. While such a dispersion relation is standard for a massive bosonic particle, it is unusual for a wave, even more so for a gravitational one; we will come back to the difference between the two.\footnote{Note that the bound on the graviton mass, recently improved by LIGO \cite{Abbott:2016blz}, does not apply to waves $h_{\mu\nu}^{{\bf k \neq 0}}$ corresponding to Kaluza--Klein modes, as mentioned in \cite{Maggiore:1900zz}. Rather, it applies to our $h_{\mu\nu}^{{\bf 0}}$, which is massless in our setup.} In our framework, we are going to argue that the mass $m$ is dominant in the dispersion relation \eqref{dispersionmassive}, leading to signals of high frequency fixed by $m$.

It is instructive to start by considering the background to be Minkowski times a flat torus $T^N$. From \eqref{finalgen} where the Riemann tensor vanishes, the $D$-dimensional gravitational wave $h_{MN}$ is then massless and can be decomposed into plane waves with $D$-dimensional light-like wave vector $k^M=(\omega,\vec{k}_3,\vec{k}_N)$, satisfying $\omega^2= \vec{k}_N^{\,2} + \vec{k}_3^{\,2}$. Looking at the four-dimensional wave $h_{\mu\nu}$ and comparing to \eqref{dispersionmassive}, one gets that $\vec{k}_3^2=\vec{p}^2$ while $m^2=\vec{k}^{\,2}_N$. The latter can also be obtained as described above \eqref{decompogen} by considering the Laplacian eigenvalue of $e^{\i \vec{k}_N \cdot \vec{y}}$. As for a standard massless wave, the vector $\vec{k}_3$ corresponds to the Minkowski spatial components of the wave vector, and is thus associated to four-dimensional physics. For instance, physics of black hole mergers in four dimensions have typical length scales leading to gravitational waves of specific wave length $\lambda_4$. The latter enters the wave vector as $|\vec{k}_3|= 2\pi/\lambda_4$. In turn, the vector $\vec{k}_N$ can be viewed as an internal momentum, and it is quantized because the torus is compact: each of its components along one circle of radius $r$ is given by an integer times $2\pi/r$. For an average radius $r_N$, we get an estimate $|\vec{k}_N|\simeq 2\pi/r_N$, which then also holds for the mass $m$. Considering more general backgrounds, these features remain true for $m$: it is quantized, because the spectrum of the Laplacian on a compact manifold is discrete,\footnote{Examples of spectra on Calabi--Yau manifolds can be found in \cite{Braun:2008jp}.} and it is given by the inverse of a typical internal length, even though the precise expression depends on the actual geometry. We conclude that the competition between $m^2$ and $\vec{p}^{\,2}$ in the dispersion relation \eqref{dispersionmassive} amounts to a comparison between the wave length $\lambda_4$, a typical four-dimensional length scale, and the typical internal length $r_N$. As we will confirm explicitly below, $r_N \ll \lambda_4$ so that $m^2 \gg \vec{p}^{\,2}$. This implies that $\omega$ is fixed by $m$, and the corresponding frequency is high compared to that associated with $\lambda_4$, i.e.~to the one of the massless wave. In addition, the quantization of $m$, related to that of the Laplacian spectrum and thus to the label ${}_{{\bf k}}$ dropped so far, leads to a discrete set of frequencies.\footnote{Such a relation between frequencies of gravitational waves and Kaluza--Klein masses was already mentioned in \cite{Clarkson:2006pq} where a specific Randall--Sundrum setup was considered (see Appendix \ref{ap:nonconstantwarp}).} In summary, there is a discrete set of extra signals $h_{\mu\nu}^{{\bf k}}$, each of them dominated by a unique angular frequency $\omega_{{\bf k}}$, which is high compared to that of the massless wave, and fixed by the Kaluza-Klein mass $m_{{\bf k}}$.

We now compute an estimate of these frequencies. We infer from above that such a frequency is given by $\nu = \omega/(2\pi) = m c^2 / (2\pi \hbar)= c/r_N$, where we restored the dependence on $c$ and $\hbar$. We then need an estimate of the typical internal length $r_N$; a review on this topic can be found in \cite{Quiros:2006my}. Current bounds from table-top experiments or missing energies in particle accelerators are $r_N \lesssim 10^{-4}\, {\rm m}$ (about $10^{-3}\, {\rm eV}$). Naively, this seems a high value for an upper bound on the size, with respect to e.g.~what would correspond to the LHC energy. But we recall that in the picture of a three-brane, presented in the Introduction, only gravity would probe the extra dimensions, while particles of the standard model would be confined to our four-dimensional space-time. Both sectors are in addition weakly coupled to each other. This value gives a lower bound for the frequency of $\nu \sim 10^{12}\, {\rm Hz}$, which can still evolve according to the considered model. The ADD models mentioned in the Introduction use $n$ extra dimensions to explain the hierarchy between the Planck mass $M_p$ and another fundamental mass scale like the electroweak one $M_*$, through the formula $(M_p/M_*)^2 = \left(M_* c / (2\pi \hbar)\right)^n r_N^n$. For $M_*= 1\, {\rm TeV.}c^{-2}$, one gets that $n=1$ extra dimension is excluded, $n=2$ corresponds approximately to the above bound, and $n=6$, as in string compactifications with three-branes, gives $r_N \sim 10^{-13}\, {\rm m}$, i.e.~a frequency $\nu \sim 10^{21}\, {\rm Hz}$. In any case, these frequency values are much higher than the typical one of the recently observed gravitational waves, around $150\, {\rm Hz}$. They are also much higher than the upper sensitivity bound of LIGO, of the order of $10^3$-$10^4\, {\rm Hz}$. In addition, future detectors seem to be planned, rather, to probe lower frequencies. This disfavors the direct detection of signals with such high frequencies, which would require a new type of apparatus. On top of being sensitive to a higher frequency range, the latter should have a much smaller strain sensitivity, to get a signal-to-noise ratio comparable to LIGO: for now this looks quite challenging. If such a detector were available, however, one could hope for a very clean signal, since there is no known astrophysical process emitting gravitational waves with frequencies much greater than $10^3\, {\rm Hz}$. Such high frequencies may thus be clear symptoms of new physics.

Let us make some final comments. There is a crucial difference between the above extra signals and a massive bosonic particle, preventing us from considering these gravitational waves as Kaluza--Klein gravitons. Classically, a particle is a localized object, that would be detected at a definite time whenever it arrives on Earth. This does not hold for waves, which can be very spread-out objects. Realistic gravitational waves are emitted continuously for a long time, by e.g.~black hole binaries getting closer to each other, increasing the wave amplitude. They are then detected only when their amplitude is high enough. This brings us to comment on the amplitude of the extra signals. A priori, we have no knowledge about it, as it depends on unknown physics. However we can still make the following remark. The energy of a gravitational wave is proportional to the square of the product of its amplitude with its frequency. Emitting signals of high frequency would then require an important energy, unless the amplitude is low. We deduce that for physical waves, the higher $\omega_{{\bf k}}$ gets, the lower the amplitude is likely to be. In principle, this does not prevent the first few modes $h_{\mu\nu}^{{\bf k}}$ from having a reasonable amplitude. Observing such a discrete set of high frequency signals, in any polarization mode, would be a very distinct signature.

\section*{Acknowledgements}

It is a pleasure to acknowledge fruitful exchanges with A.~Buonanno, X.~Calmet, E.~Dudas, H.~Godazgar, A. Harte, M.~Henneaux, J.~Korovins, S. Marsat, K.~Mkrtchyan, R. Rahman, V.~Raymond, C.~Troessaert and D.~Tsimpis. The research of G\,LG was partially supported by the Alexander von Humboldt Foundation. Our warmest thanks go to the Deutsche Bahn AG for providing us with comfortable office space in their conveniently delayed trains.

\newpage

\begin{appendix}

\section{Further analysis of the equations}

\subsection{Equations in the Transverse-Traceless gauge}\label{ap:TT}

In the main text, we have worked with the $D$-dimensional de Donder gauge \eqref{Dondergauge}. For completeness, we present here a standard refinement of the latter, namely the $D$-dimensional Transverse-Traceless (TT) gauge:
\beq
g^{PQ} \nabla_P^{(0)} h_{QN} = 0 \ ,\quad h_D = 0 \ .\label{TTgaugechoice}
\eeq
To implement this $D$-dimensional gauge in the three wave equations \eqref{eqgen4d}, \eqref{eqgenoffdiag} and \eqref{eqgenintern}, we first compute, using formulas of Section \ref{sec:split},
\begin{subequations}
\label{TTDdimensionsnonconstantwarp}
\begin{align}
h_D = 0\quad &\Leftrightarrow\quad e^{-2A} \tilde{h}_4 + h_N  = 0 \,,\label{TT1}\\
g^{PQ} \nabla_P^{(0)} h_{Q\nu} = 0\quad &\Leftrightarrow\quad e^{-2A} \tilde{g}^{\pi\rho} \tilde{\nabla}_{\pi} h_{\rho \nu} + g^{pq} \nabla_p h_{q\nu} + 2 h_{\nu q} g^{qp} e^{-2A} \del_p e^{2A} = 0 \,,\label{TT2nonconstantwarp}\\
g^{PQ} \nabla_P^{(0)} h_{Qr} = 0\quad &\Leftrightarrow\quad e^{-2A} \tilde{g}^{\pi\rho} \tilde{\nabla}_{\pi} h_{\rho r} + 2 e^{-2A} h_{rq} g^{qp} \del_p e^{2A} - \frac{1}{2} \tilde{h}_4 e^{-4A} \del_r e^{2A} \label{TT3nonconstantwarp}\\
& \hspace*{24pt} + g^{pq} \nabla_p h_{qr} = 0 \,.\nn
\end{align}
\end{subequations}
Those conditions are the analogue to the de Donder ones \eqref{DD1} and \eqref{DD2}. The three wave equations then become, in the $D$-dimensional TT gauge,
\begin{subequations}
	\begin{align}
{}_{\mu\nu}:\ & e^{-2A} \tilde{\square}_4 h_{\mu\nu} + \Delta_{\mmm} h_{\mu\nu} - h_{\mu\nu} \Delta_{\mmm} \ln e^{2A} \label{eqgen4dTT}\\
& -2 \tilde{{\cal R}}^{\pi}{}_{\mu\nu\sigma} g^{\sigma \rho} h_{\rho \pi} - \frac{1}{2} e^{-2A} g^{pq} \del_p e^{2A} \del_q e^{2A} \left( \tilde{g}_{\nu\mu} h_4 +2 h_{\nu\mu} e^{-2A} \right) \nn\\
& +2 e^{-2A} \tilde{\nabla}_{(\mu} h_{\nu) m} g^{mn} \del_n e^{2A} - \tilde{g}_{\mu\nu} h_{mn} g^{mr} g^{np} \left( \nabla_r \del_p e^{2A} - e^{-2A}\del_r e^{2A} \del_p e^{2A} \right) = 0\,, \nn\\
{}_{\mu n}:\ & e^{-2A} \tilde{\square}_4 h_{\mu n} + \Delta_{\mmm} h_{\mu n} + e^{-2A} g^{pq} \nabla_p h_{\mu n} \del_q e^{2A} + e^{-2A} g^{pq} \nabla_p h_{q\mu} \del_n e^{2A} \label{eqgenoffdiagTT}\\
& + e^{-2A} h_{\mu m} g^{mp} \nabla_n\del_p e^{2A}  - e^{-4A} h_{\mu n} g^{pq} \del_p e^{2A} \del_q e^{2A} - \frac{1}{2} h_{\mu n} \Delta_{\mmm} \ln e^{2A} \nn\\
& + e^{-2A} g^{pq} \del_{\mu} h_{np} \del_q e^{2A} = 0 \,,\nn\\
{}_{m n}:\ & e^{-2A} \tilde{\square}_4 h_{mn} + \Delta_{\mmm} h_{mn} + 2 e^{-2A} g^{pq} \del_p e^{2A} \nabla_q h_{mn} + 2 e^{-4A} g^{pq} \del_p e^{2A} h_{q(m} \del_{n)} e^{2A}  \label{eqgeninternTT}\\
 &\!\!\! -2 {\cal R}^s{}_{mnp} g^{pq} h_{qs} + 2 e^{-2A} g^{pq} \nabla_{p} h_{q (m} \del_{n)} e^{2A} + e^{-2A}  h_N \nabla_n \del_m e^{2A} = 0 \,.\nn
	\end{align}
\end{subequations}
The four-dimensional component $h_{\mu\nu}$ is not present in the two equations \eqref{eqgenoffdiagTT} and \eqref{eqgeninternTT}. The dynamics of $h_{\mu n}$ and $h_{m n}$ thus seem to decouple; however, one should keep in mind the coupling present through the gauge conditions \eqref{TT1}, \eqref{TT2nonconstantwarp} and \eqref{TT3nonconstantwarp}. Still, the contributions of $h_{\mu n}$ and $h_{m n}$ can be viewed as source terms in the equations describing the dynamics of $h_{\mu\nu}$. As a side remark, note that further constraints on $h_{\mu n}$ and $h_{m n}$ seem obtainable by tracing \eqref{eqgen4dTT}.

\subsection{The case of a non-constant warp factor}\label{ap:nonconstantwarp}

In the main text, we have considered the case of a constant warp factor. Let us come back here to the study of the wave equations \eqref{eqgen4d}, \eqref{eqgenoffdiag} and \eqref{eqgenintern}, obtained in the $D$-dimensional de Donder gauge, without assuming a constant warp factor. They display an intricate mixing of $h_{\mu\nu}$ with the vector and scalar components, making the system of equations involved. We are eventually interested in the four-dimensional dynamics, and hence in the following we focus on a particular type of solutions, that is,
\beq
h_{\mu n}= 0 \ ,\ h_{mn}=0 \ . \label{h=0}
\eeq
On top of the simplification, there are two reasons for considering \eqref{h=0}. Firstly, imposing this condition, the two equations of motion \eqref{eqgenoffdiag} and \eqref{eqgenintern} boil down to
\beal
\tilde{g}^{\pi \rho} \tilde{\nabla}_{\pi} h_{\rho \nu}\ \del_n e^{2A} &= 0 \ ,\\
\tilde{h}_4\ \nabla_n \del_m \ln e^{2A} &= 0 \ .
\eeal
Tracing the last equation yields $\Delta_{\mmm} \ln e^{2A}$, and on a compact manifold this Laplacian vanishes if and only if $\ln e^{2A}$ is constant. Considering a non-constant warp factor then makes these equations equivalent to
\beq
\tilde{g}^{\pi \rho} \tilde{\nabla}_{\pi} h_{\rho \nu} = 0\ ,\quad \tilde{h}_4 = 0 \ , \label{4dgaugeTT}
\eeq
i.e.~the four-dimensional TT gauge. This is here equivalent to the $D$-dimensional TT gauge \eqref{TTgaugechoice}, as the latter reduces to \eqref{4dgaugeTT} when imposing \eqref{h=0}. Remarkably, this TT gauge fixing is enforced by the two extra equations of motion, while only the $D$-dimensional de Donder gauge had been imposed.\footnote{Suppose we do not impose the $D$-dimensional de Donder gauge, and keep the corresponding terms as in \eqref{final1}. Imposing then \eqref{h=0}, the ${}_{\mu n}$ and ${}_{mn}$ components of the first order Einstein equation do not give any relevant condition: not a TT gauge nor a de Donder gauge.} Eventually, we are left with only the four-dimensional equation \eqref{eqgen4d}.

Secondly, imposing~\eqref{h=0} means that if one starts, rather, with equations \eqref{eqgen4dTT}, \eqref{eqgenoffdiagTT} and \eqref{eqgeninternTT}, where the $D$-dimensional TT gauge has been imposed, then $h_{\mu n}$ and $h_{mn}$ act as sources for $h_{\mu\nu}$, as pointed-out in Appendix \ref{ap:TT}. Imposing \eqref{h=0} then amounts to considering the four-dimensional equation \eqref{eqgen4dTT} without sources, which is a legitimate way to proceed; consistently, the two equations \eqref{eqgenoffdiagTT} and \eqref{eqgeninternTT} vanish when imposing the condition \eqref{h=0}.\\

In either case, one is left only with the study of the following four-dimensional equation:
\beq
e^{-2A} \tilde{\square}_4 h_{\mu\nu} + \Delta_{\mmm} h_{\mu\nu} - h_{\mu\nu} \left( \Delta_{\mmm} \ln e^{2A} + e^{-4A} ( \del e^{2A} )^2 + \frac{2}{3} e^{-2A} \Lambda_4 \right) = 0 \ , \label{eqh}
\eeq
where $( \del e^{2A} )^2 = g^{pq} \del_p e^{2A} \del_q e^{2A}$ and \eqref{bla} has been used. Furthermore, using the relation $\Delta_{\mmm} \ln e^{2A} = e^{-2A} \Delta_{\mmm} e^{2A} - e^{-4A} ( \del e^{2A} )^2 $ and redefining the field as $\tilde{h}_{\mu\nu} = e^{-2A} {h}_{\mu\nu}$, the perturbation with respect to $\tilde{g}_{\mu\nu}$ instead of $g_{\mu\nu}$, the above equation reads
\beq
 \tilde{\square}_4 \tilde{h}_{\mu\nu} + e^{2A} \Delta_{\mmm} \tilde{h}_{\mu\nu} + 2 g^{pq} \del_p \tilde{h}_{\mu\nu} \del_q e^{2A} - \frac{2}{3} \Lambda_4 \tilde{h}_{\mu\nu} = 0 \ . \label{eqtildeh}
\eeq
Interestingly, this equation matches precisely an equation obtained in \cite{Davoudiasl:1999jd}, where the RS1 model \cite{Randall:1999ee} discussed in the Introduction was taken as the background. This is a particular case of the present framework: indeed, the RS1 background fits our metric \eqref{backgroundgen}, with a four-dimensional Minkowski space-time, $N=1$ extra dimension, and an explicit non-constant warp factor, where $A$ is linear in the extra coordinate. Such a warp factor is unusual with respect to standard supergravity expressions and brane solutions, but it remains appealing as a phenomenological model. This matching with \cite{Davoudiasl:1999jd} is an important cross-check, and indicates possible generalisations beyond the RS1 model.

Rewritings of equation \eqref{eqtildeh} can be obtained by considering two different rescalings. First, the internal metric can be rescaled as $g_{mn}=e^{-2A} \tilde{g}_{mn}$, with the following motivation. The warp factor typically accounts for the backreaction of an extended object such as a $D_p$-brane, often present in models with extra dimensions. The $D_p$-branes also correspond to solutions of the equations of motion of ten-dimensional type II supergravities. So far, the background equations were assumed to be satisfied: with a non-constant warp factor, those would be solved by a $D_p$-brane solution. Thus we consider a $D_3$-brane-like background solution, where the brane fills the three external space dimensions, i.e.~is transverse to $\mmm$. This solution requires extracting the warp factor as $g_{mn}=e^{-2A} \tilde{g}_{mn}$~\cite{Giddings:2001yu}. Note that this solution also requires ingredients beyond $D$-dimensional gravity, namely a $5$-form flux $F_5^{10}$, that contributes to the Einstein equation. Such contributions are studied in Appendix \ref{ap:gen}. For now, we simply rescale the internal metric. The two corresponding internal Laplacians on a scalar $\varphi$ are then related by $\Delta_{\mmm} \varphi = e^{2A} \tilde{\Delta}_{\mmm} \varphi + \frac{1}{2} (6 - D) \tilde{g}^{pq} \del_p e^{2A} \del_q \varphi$. Then, \eqref{eqtildeh} becomes
\beq
 \tilde{\square}_4 \tilde{h}_{\mu\nu} + e^{4A} {\tilde \Delta}_{\mmm} \tilde{h}_{\mu\nu} + \frac{10 - D}{2} e^{2A} \tilde{g}^{pq} \del_p \tilde{h}_{\mu\nu} \del_q e^{2A} - \frac{2}{3} \Lambda_4 \tilde{h}_{\mu\nu} = 0 \ . \label{eqtildehtildeg}
\eeq
This equation would be interesting in the context of type II supergravities where $D=10$. Finally, another rescaling consists in having the same warp factor over all dimensions, leading us to introduce $g_{mn}=e^{2A} \bar{g}_{mn}$. Using the above Laplacian formula reversewise, one rewrites \eqref{eqtildeh} to
\beq
 \tilde{\square}_4 \tilde{h}_{\mu\nu} + \bar{\Delta}_{\mmm} \tilde{h}_{\mu\nu} + \frac{D-2}{2} e^{-2A} \bar{g}^{pq} \del_p \tilde{h}_{\mu\nu} \del_q e^{2A} - \frac{2}{3} \Lambda_4 \tilde{h}_{\mu\nu} = 0 \ . \label{eqtildehbarg}
\eeq
This matches precisely the equation obtained on general grounds in \cite{Bachas:2011xa} with $h_{mn}=h_{\mu n}=0$ in 4d TT gauge. That work also generalises \cite{Csaki:2000fc} where the same analysis was carried-out with only one extra dimension. Here again, we obtain an important computational cross-check, and foresee possible generalisations of previous works.

Whether one considers \eqref{eqh}, \eqref{eqtildeh}, \eqref{eqtildehtildeg} or \eqref{eqtildehbarg}, the next step of the analysis would require information on the extra dimensions. Restricting to $\mmm$ being compact, one can expand $\tilde{h}_{\mu\nu} (x,y)$ as an infinite sum of Kaluza--Klein modes $ \tilde{h}_{\mu\nu}^{{\bf k}}(x)$ on an internal basis, as e.g.~in~\eqref{decompogen}. This leads to a set of equations for these various modes. The difficulty is to define this basis in such a way that all the modes of the four-dimensional wave decouple, each one of them described by independent equations, as in the case of a constant warp factor in Section \ref{sec:constantwarp}. In particular the zero-mode should correspond to a massless four-dimensional gravitational wave. Here, in the case of a non-constant warp factor, one first builds a differential operator acting on $\tilde{h}_{\mu\nu} (x,y)$ by combining the two terms with internal first and second derivatives in \eqref{eqtildeh}, \eqref{eqtildehtildeg} or \eqref{eqtildehbarg}. The appropriate basis is then that of eigenmodes of this operator. We refer e.g.~to \cite{Bachas:2011xa} for more detail. The same procedure was used in phenomenological papers working with the RS1 background. Motivated by a detection of Kaluza--Klein gravitons at the LHC, this allowed to determine explicitly the Kaluza--Klein mass spectrum on such a background in \cite{Davoudiasl:1999jd} (see \cite{Dillon:2016fgw} for a recent presentation and further references).\footnote{Such an analysis requires to define a mass $m$ on a maximally symmetric space-time: we define it as $\tilde{\square}_4 \tilde{h}_{\mu\nu}  - \frac{2}{3} \Lambda_4 \tilde{h}_{\mu\nu}= m^2 \, \tilde{h}_{\mu\nu}$. This unphysical mass definition allows unitary scalars on an anti-de Sitter space-time to have negative $m^2$ by the Breitenlohner--Freedman bound (an analogue on de Sitter is discussed in \cite{Strominger:2001pn, McInnes:2001dq}). However, it should be noted that it is not the case for massive spin-$2$ fields, which must have strictly positive $m^2$, see e.g.~the review \cite{Rahman:2015pzl}.} In a similar context, another work \cite{Clarkson:2006pq} determined such a Kaluza--Klein spectrum numerically and related it to frequencies of gravitational waves. More generally, getting concrete results from this procedure requires some knowledge of the warp factor, as it enters the eigenmode equation; but this is typically not the case in string compactifications. We still hope to come back to this case of a non-constant warp factor, which could lead to further interesting effects of extra dimensions on the four-dimensional gravitational waves.

\section{Completing the model}\label{ap:gen}

Models with extra dimensions often contain more than $D$-dimensional General Relativity with a cosmological constant. In this appendix, we work-out the contributions of this additional content. We start by considering the following general action:
\beq
S= \frac{1}{2 \kappa_D} \int \d^{D} x \, \sqrt{|g_{D}|}\ \left( {\cal R}_{D} + L  \right)\ , \label{actiongen}
\eeq
where $L$ generically depends on $g_{D\ MN}$, without containing any derivative of this metric. This captures the previous \eqref{action} with a cosmological constant, but also e.g.~a Yang--Mills term or more generally standard supergravities in Einstein frame, recalling that field strengths are fully antisymmetrized tensors in which derivatives of the metric are absent thanks to the Levi--Civita connection. We introduce the quantities
\beq
L_{MN}= \frac{\del L}{\del g_D^{MN}} \ , \ L_D= g_D^{MN} L_{MN} \ .
\eeq
The no-derivative property allows to derive the Einstein equation without integration by parts, giving
\beq
{\cal R}_{MN} + L_{MN} - \frac{g_{D\ MN}}{2}\left({\cal R}_D + L \right) = 0 \ . \label{EinsteinDgen}
\eeq
As in Section \ref{sec:firstderiv}, we use the trace
\beq
{\cal R}_D = \frac{2D}{D-2} \left( \frac{L_D}{D} - \frac{L}{2} \right)\ , \label{Einsteintracegen}
\eeq
to rewrite the Einstein equation as
\beq
{\cal R}_{MN} + L_{MN} - \frac{g_{D\ MN}}{D-2}\left( L_D - L \right) = 0 \ .\label{EinsteinL}
\eeq
We proceed further as in Section \ref{sec:firstderiv} and write the above at zeroth and first order in $h_{MN}$,
\bea
& {\cal R}^{(0)}_{MN} + L^{(0)}_{MN} - \frac{g_{MN}}{D-2}\left( L^{(0)}_D - L^{(0)} \right) = 0 \ ,\label{EinsteinL0}\\
& {\cal R}^{(1)}_{MN} + L^{(1)}_{MN} - \frac{g_{MN}}{D-2}\left( L^{(1)}_D - L^{(1)} \right) - \frac{h_{MN}}{D-2}\left( L^{(0)}_D - L^{(0)} \right) = 0 \ , \label{EinsteinL1}
\eea
where
\beq
L^{(1)}_D= g^{MN} L^{(1)}_{MN} - g^{PQ} h_{QR} g^{RS} L^{(0)}_{PS} \ .
\eeq
We now read the Ricci tensor at first order from \eqref{Ricci2}. There, we replace ${\cal R}^{(0)}_{MP}$ using the background Einstein equation \eqref{EinsteinL0}. The first order Einstein equation \eqref{EinsteinL1} becomes
\bea
- \frac{1}{2} \square_D^{(0)}\, h_{MN}  + {\cal R}^{(0)}{}^S{}_{MNP} &\,  g^{PQ} h_{QS} + \nabla^{(0)}_{(M} {\cal G}_{N)} \label{EinsteinL1bis} \\
&  = \frac{g_{MN}}{D-2} \left(  L^{(1)}_D - L^{(1)} \right) -  L^{(1)}_{MN}  + g^{PQ} h_{Q(N} L^{(0)}_{M)P} \,.\nn
\eea
Finally, the $D$-dimensional de Donder gauge \eqref{Dondergauge}, to be considered from now on, makes the ${\cal G}_{N}$ term disappear.

\subsection*{Constant warp factor}

Let us specify here as in Section \ref{sec:constantwarp} to the case of a constant warp factor: we set $A=0$ and drop the tilde. The components ${}_{\mu\nu}$ and ${}_{mn}$ of \eqref{EinsteinL1bis} are then given by
\begin{subequations}
	\begin{align}
\hspace{-0.2cm} -\frac{1}{2} \square_4 h_{\mu\nu} -\frac{1}{2} \Delta_{\mmm} h_{\mu\nu} + {\cal R}^{\pi}{}_{\mu\nu\sigma} g^{\sigma \rho} h_{\rho \pi} & = \frac{g_{\mu\nu}}{D-2} \left(  L^{(1)}_D - L^{(1)} \right) -  L^{(1)}_{\mu\nu}  + g^{\pi\rho} h_{\rho(\nu} L^{(0)}_{\mu)\pi} \, , \label{constantwarp4dL} \\
\hspace{-0.2cm} -\frac{1}{2} \square_4 h_{mn} -\frac{1}{2} \Delta_{\mmm} h_{mn} + {\cal R}^s{}_{mnp} g^{pq} h_{qs} & = \frac{g_{mn}}{D-2} \left(  L^{(1)}_D - L^{(1)} \right) -  L^{(1)}_{mn}  + g^{pq} h_{q(n} L^{(0)}_{m)p} \, , \hspace*{-5pt}\label{constantwarpNdL}
	\end{align}
\end{subequations}
using that $L^{(0)}_{\mu n} = 0$. Let us now consider the trace of \eqref{constantwarpNdL}. It involves the Ricci tensor ${\cal R}^{(0)}_{mn}$: the latter can be replaced using the background equation of motion \eqref{EinsteinL0}, leading to
\beq
-\frac{1}{2} \square_4 h_N -\frac{1}{2} \Delta_{\mmm} h_N =  - g^{mn} L^{(1)}_{mn} + \frac{N}{D-2} \left(  L^{(1)}_D - L^{(1)} \right) + \frac{h_N}{D-2} \left(  L^{(0)}_D - L^{(0)} \right) \ . \label{constantwarpNdLtrace}
\eeq
In Section \ref{sec:constantwarp}, considering a Minkowski background forced us to restrict to a Ricci flat internal manifold. Here, the additional content allows us to consider more general Minkowski backgrounds: see e.g.~\cite{Andriot:2015sia} and references therein for explicit examples, and \cite{Andriot:2016ufg} for a whole class of such backgrounds of type II supergravities. To pursue the same reasoning and get similar effects as in Section \ref{sec:constantwarp}, despite a more general background, a simple condition would be to have the same starting equations. This implies that the right-hand sides of \eqref{constantwarp4dL} and \eqref{constantwarpNdLtrace} should vanish. It would be interesting to check whether these two conditions hold, at least for some example. In addition, note that a Minkowski background implies through \eqref{EinsteinL0} that $L^{(0)}_{\mu\nu} = \tfrac{g_{\mu\nu}}{D-2} ( L^{(0)}_D - L^{(0)} )$.

\subsection*{General remarks and non-constant warp factor}

We now make general remarks on the new contributions to the four-dimensional equation. As an illustration for those, we focus on the quantity $L_{MN=\mu\nu}$, that gives rise to the term $-  L^{(1)}_{\mu\nu}  + g^{PQ} h_{Q(\nu} L^{(0)}_{\mu)P}$ in \eqref{EinsteinL1bis}. A further motivation to analyse the latter is that it could provide additional interesting terms in the case of a non-constant warp factor, as mentioned in Appendix \ref{ap:nonconstantwarp}.

The background quantities should preserve four-dimensional Lorentz invariance, since we consider an empty space-time. As a consequence, the only non-trivial four-dimensional tensors are the four-dimensional metric, its volume form, and derivatives thereof. $L_{MN}$ does not contain derivatives of the metric, so that $L^{(0)}_{MN=\mu\nu}$ must be proportional to $g_{\mu\nu}$. In addition, the only possible four-dimensional background fluxes must be forms of degree $4$ (or higher), proportional to the four-dimensional volume form. For instance, if $L$ is a standard abelian Yang--Mills term, $L^{(0)}_{MN=\mu\nu}$ vanishes, because the four-dimensional $2$-form field strength has to vanish. We will make use of this general characterization.

We now consider the $D$-dimensional model to be a ten-dimensional type II supergravity; for conventions, we refer to \cite{Andriot:2016xvq}. A reason to do so is that this framework provides background solutions with non-constant warp factor. We will eventually focus on one of them, the $D_3$-brane. As explained in Appendix \ref{ap:nonconstantwarp}, having a non-constant warp factor indicates the presence of an extended object such as a brane, which in turn sources a flux. Type II supergravities contain all such ingredients on top of gravity, and are thus suitable models here. Let us first look at contributions from the $D_p$-branes and orientifolds $O_p$-planes. Those are $p$-dimensional extended objects, here along the three space directions of the four-dimensional space-time and $p-3$ internal dimensions; this property prevents them from breaking four-dimensional Lorentz invariance. The part of their action that contributes to the Einstein equation is the Dirac--Born--Infeld (DBI) action. With few assumptions, see e.g.~\cite{Andriot:2016xvq}, it can be written as
\beq
S_{DBI} = {\rm constant} \times \int \d^{10}x\ e^{-\phi} \sqrt{|g_{10}|} \ \frac{ \delta(\bot) }{\sqrt{|g_{\bot}|}} \ ,
\eeq
where $\phi$ is the dilaton scalar field, $g_{\bot}$ denotes the metric along the internal transverse directions to the object and $\delta(\bot)$ localizes it in these directions. As a consequence, $L_{MN=\mu\nu} = 0$ for these ingredients.

The background solution of interest includes $D_3$-branes and $O_3$-planes. For those, the dilaton is constant and given by $e^{\phi} = g_s$. This way, it does not contribute to the Einstein equation, and one avoids complications related to the string versus Einstein frame. The only contributions left to study are thus those of fluxes. They enter $L_{MN=\mu\nu}$ through their components having at least one leg along the four dimensions. As explained previously, this restricts them to be a form of degree $4$ or higher. In type IIB supergravity with $D_3/O_3$, the only appropriate flux is a $5$-form denoted $F^{10}_5$. This flux is a crucial ingredient of the background solution, as being sourced by the $D_3/O_3$. It contributes as follows
\beq
L= -\frac{g_s^{2}}{4}  |F_5^{10}|^2 \ , \quad L_{MN}= - \frac{g_s^{2}}{4\cdot4!} F_{5\ MPQRS}^{10}F_{5\ N}^{10 \ \ PQRS} \ ,
\eeq
where for a $p$-form $A_p$, we denote $|A_p|^2 = A_{p\, M_1 \dots M_p}  A_p^{M_1 \dots M_p} / p!$, raising indices with $g_{D\, MN}$. Its background components are either proportional to the four-dimensional volume form, or purely internal: $F_5^{10\, (0)}= F_5^4 + F_5$, where $F_5^4= g_s^{-1}\, {\rm vol}_4 \w f_5$ with an internal $1$-form $f_5$. In the actual solution, $f_5= e^{-4A} \d e^{4A}= - g_s *_6 F_5$, even though we will not use this expression. We deduce
\beq
L^{(0)}_{MN=\mu\nu}= - \frac{|g_4|}{4\cdot3!}   \varepsilon_{\mu\rho\pi\sigma} \varepsilon_{\nu}^{\ \ \rho\pi\sigma} |f_5|^2 = - \frac{|g_4|}{4}  \frac{g_{\mu\nu}}{g_4} |f_5|^2 =  \frac{1}{4} g_{\mu\nu} |f_5|^2 \ ,
\eeq
where indices are raised with $g_{\kappa \lambda}$.

Finally, we evaluate $L^{(1)}_{MN=\mu\nu}$, considering only first order fluctuations of the metric,  meaning that $F_5^{10}$ remains at its background value. To connect to Appendix \ref{ap:nonconstantwarp}, we also set from now on $h_{\mu m}= h_{mn}=0$ as in \eqref{h=0}, and get
\beq
L^{(1)}_{MN=\mu\nu}=  \frac{|g_4|}{4\cdot2!} \varepsilon_{\mu\rho\pi\sigma} \varepsilon_{\nu\tau}^{\ \ \ \pi\sigma} g^{\rho \kappa} h_{\kappa \lambda} g^{\lambda \tau}  |f_5|^2 \ .
\eeq
Since $\tfrac{1}{2}\varepsilon_{\mu\rho\pi\sigma} \varepsilon_{\nu\tau}^{\ \ \ \pi\sigma} = \tfrac{1}{g_4}\left( g_{\mu\nu} g_{\rho \tau} - g_{\mu\tau} g_{\rho \nu} \right)$, we conclude
\beq
L^{(1)}_{MN=\mu\nu}=  -\frac{1}{4} \left( g_{\mu\nu} h_4 - h_{\mu \nu} \right) |f_5|^2 \ .
\eeq
We finally compute the relevant combination in \eqref{EinsteinL1bis} in four dimensions and obtain
\beq
 -  L^{(1)}_{\mu\nu}  + g^{PQ} h_{Q(\nu} L^{(0)}_{\mu)P} = \frac{1}{4} g_{\mu\nu}\, h_4 \ .\label{contribF5}
\eeq
The terms in $h_{\mu\nu}$ have been canceled! In this framework and with such a background, we thus conclude that the additional content does not provide any four-dimensional mass term, i.e.~a term proportional to $h_{\mu \nu}$. In addition, given the above discussion, it seems to be a fairly general result; in particular, a similar result holds for an $F_4$ flux.

For completeness, we considered as well first order fluctuations of the flux. For flux fluctuations to contribute to $L^{(1)}_{MN=\mu\nu}$ in the form of a mass term, they should produce $h_{\mu\nu}$, meaning be equal in value, as a solution to the first order equation. Despite interesting possibilities from the flux components partially along four dimensions and along $\mmm$, with a flux fluctuation weighting the square root of the $h_{\mu\nu}$ fluctuation, we did not find any satisfying solution. Flux fluctuations are subject to the flux equation of motion and Bianchi identity, which are too constraining. It would be interesting to analyse further the new contributions to the Einstein equation.

\end{appendix}

\newpage

\providecommand{\href}[2]{#2}\begingroup\raggedright

\endgroup

\end{document}